\documentclass[12pt]{article}
\usepackage{latexsym}
\usepackage{amsfonts}
\usepackage{amssymb}
\usepackage{graphicx}
\usepackage{color}
\begin{document}
\input epsf
\def\be{\begin{equation}}
\def\bea{\begin{eqnarray}}
\def\ee{\end{equation}}
\def\eea{\end{eqnarray}}
\def\d{\partial}
\definecolor{red}{rgb}{1,0,0}
\long\def\symbolfootnote[#1]#2{\begingroup%
\def\thefootnote{\fnsymbol{footnote}}\footnote[#1]{#2}\endgroup}
\renewcommand{\a}{\left( 1- \frac{2M}{r} \right)}
\newcommand{\dm}{\begin{displaymath}}
\newcommand{\edm}{\end{displaymath}}
\newcommand{\com}[2]{\ensuremath{\left[ #1,#2\right]}}
\newcommand{\la}{\lambda}
\newcommand{\eps}{\ensuremath{\epsilon}}
\newcommand{\half}{\frac{1}{2}}
\newcommand{\field}[1]{\ensuremath{\mathbb{#1}}}
\renewcommand{\l}{\ell}
\newcommand{\bl}{\left(\l\,\right)}
\newcommand{\normljk}{\langle\l,j,k|\l,j,k\rangle}
\newcommand{\N}{\mathcal{N}}
\renewcommand{\b}[1]{\mathbf{#1}}
\renewcommand{\v}{\xi}
\newcommand{\tr}{\tilde{r}}
\newcommand{\ttheta}{\tilde{\theta}}
\newcommand{\tgamma}{\tilde{\gamma}}
\newcommand{\bg}{\bar{g}}
\newcommand{\implies}{\Rightarrow}
\newcommand{\z}{\ensuremath{\ell_{0}}}
\newcommand{\temp}{\ensuremath{\sqrt{\frac{2\z+1}{\z}}}}
\newcommand{\twomatrix}[4]{\ensuremath{\left(\begin{array}{cc} #1 & #2
\\ #3 & #4 \end{array}\right) }}
\newcommand{\columnvec}[2]{\ensuremath{\left(\begin{array}{c} #1 \\ #2
\end{array}\right) }}
\newcommand{\e}{\mbox{\textbf{e}}}
\newcommand{\gm}{\Gamma}
\newcommand{\bt}{\bar{t}}
\newcommand{\bphi}{\bar{\phi}}
\newcommand{\m}{\ensuremath{\mathbf{m}}}
\newcommand{\n}{\ensuremath{\mathbf{n}}}
\renewcommand{\theequation}{\arabic{section}.\arabic{equation}}
\newcommand{\newsection}[1]{\section{#1} \setcounter{equation}{0}}

\vspace{20mm}
\begin{center} {\LARGE Smooth geometries with four charges in four dimensions }
\\
\vspace{20mm} {\bf  Ashish Saxena$^{1}$, Geoff Potvin$^{1}$, Stefano Giusto$^{2}$ and Amanda W. Peet$^{1}$}\\ 
\vspace{2mm}
\symbolfootnote[0]{ {\tt ashish@physics.utoronto.ca, gpotvin@physics.utoronto.ca, giusto@mps.ohio-state.edu} and {\tt amanda.peet@utoronto.ca}} 
$^{1}$Department of Physics,\\ University of Toronto,\\ 
Toronto, Ontario, Canada M5S 1A7;\\
\vspace{2mm}
and\\
\vspace{2mm}
$^{2}$Department of Physics,\\ The Ohio State University,\\ 
Columbus, Ohio, USA 43210.\\
\vspace{4mm}
\end{center}
\vspace{10mm}
\begin{abstract}
A class of axially symmetric, rotating four-dimensional geometries carrying D1, D5, KK monopole and momentum charges is constructed. The geometries are found to be free of horizons and singularities, and are candidates to be the gravity duals of microstates of the (0,4) CFT. These geometries are constructed by performing singularity analysis on a suitably chosen class of solutions of six-dimensional minimal supergravity written over a Gibbons-Hawking base metric. The properties of the solutions raise some interesting questions regarding the CFT. 
\end{abstract}

\thispagestyle{empty}
\newpage
\setcounter{page}{1}

\newsection{Introduction}\setcounter{equation}{0}

The AdS/CFT duality \cite{Maldacena,Witten} has been a major driving force behind our enhanced understanding of nonperturbative aspects of string theory. Particularly important for issues relating to black holes has been the D1-D5 system \cite{SpentaReview}. Wrapping a large number of parallel D1 and D5 branes on a five torus and adding momentum along the D1 branes produces the classic example of a five dimensional black hole in string theory with non-zero entropy and a regular horizon \cite{bmpv}. This system has been extensively analysed both from the supergravity \cite{cy} and CFT \cite{Larsen5D} points of view, and is therefore ideal for asking questions related to black hole microscopics. For example, the Bekenstein-Hawking entropy can be completely accounted for by counting relevant states in the world volume theory on the branes~\cite{stromvafa}. However, the $AdS_3/CFT_2$ duality raises the prospect of being able to see the states responsible for the entropy not only in the associated field theory but also on the gravity side. To be precise, a general state in the CFT would be dual to some state in the full quantum string theory on $AdS_3$ (which may or may not be well approximated by a gravity solution). For large enough $N$, one can imagine building a ``coherent'' state in the CFT which would be sufficiently classical so as to admit a supergravity description on the string side. As far as AdS/CFT is concerned, the gravity dual would be an asymptotically AdS geometry. To regard this state as a microstate of an asymptotically flat black hole one would have to somehow continue this geometry so as to add an asymptotically flat region. The Mathur conjecture~\cite{mss,lmstat} is a concrete proposal in this direction.  The idea is to associate the above-mentioned coherent state in the CFT with an asymptotically flat geometry which is smooth, free of horizons, carries the same conserved quantum numbers as the associated black hole, and hence constitutes a microstate of the relevant black hole. 

Thus far, support for the conjecture comes primarily from studying the D1-D5 system. The strongest piece of evidence is the complete mapping of the ground states of the D1-D5 system in the Ramond sector by Lunin and Mathur~\cite{lmadscft}. As an added bonus one also finds the gravity duals of chiral primaries in the NS sector of the CFT~\cite{lms}. The Lunin-Mathur geometries do not have horizons and are either completely smooth or have acceptable orbifold singularities which are resolved in the full string theory \cite{lmm}. The amenability of this class lies in the fact that the D1-D5 system compactified on a five torus can be mapped to a fundamental string carrying winding and momentum. This picture also allows one to make precise the notion of coherent states mentioned above. For a recent review of these issues, see \cite{MathurRecentReview}.  

The situation becomes much more complex when one adds momentum to the system. This configuration cannot be mapped to anything simpler by a chain of dualities and it is not known how to attack the problem for arbitrary amounts of momentum. For some simple states, however, gravity duals have been found~\cite{gms1,gms2,lunin1}. These geometries exhibit the same basic features as the D1-D5 states -- they are free of horizons and singularities. In~\cite{gms1}, the solution was derived by taking the extremal limit of a class of non-extremal three charge black hole solutions \cite{cy}. Generically these solutions have singularities and horizons. However, for the very special values which arise from the CFT the geometries are completely smooth. Amazingly, it was later found that there were non-BPS microstates with CFT interpretation hiding in this large class of solutions which exhibit the same smoothness as the BPS states found earlier\cite{simon1}.

Given the progress in testing the conjecture for five dimensional black holes, it is very natural to ask whether this success can be extended to four dimensions. After all, nature seems to be rather partial to four dimensions. One can easily construct brane configurations in string theory which lead to four dimensional black holes. Particularly suited to this discussion is the D1-D5-KK6 system\cite{Tseytlin1,Tseytlin2}. Unlike the D1-D5 case, however, much less is known about the CFT underlying this system \cite{Larsen4D}. Furthermore, until recently \cite{blackhole4D} it seemed impossible to construct a four dimensional rotating BPS black hole.  Though one can write down a rotating non-extremal class of solutions carrying four charges, in the BPS limit the presence of angular momentum leads to a naked singularity similar to the zero mass (but non-zero angular momentum) Kerr solution. The experience with the corresponding five dimensional solutions might make one wonder whether there might be isolated values of parameters for which the singularities are resolved and one is left with a horizon-free solution. As shown in Appendix A, there are no solutions in the BPS limit of this class which are horizon- and singularity-free. It does not seem that there are any smooth non-BPS states either\footnote{We would like to thank Simon Ross and Omid Saremi for discussion on this point.}. 
  
The work of Bena and Kraus in~\cite{bk} changed this rather dismal situation by directly constructing a rotating \emph{and} supersymmetric solution carrying D1, D5 and KK monopole charges. This was made possible by the recent advances in our understanding of supersymmetric solutions in various supergravities\cite{bw,gmr}. Their solution relied on the previous work of~\cite{bw} and was constructed by performing a singularity analysis on a suitable class of explicitly known supersymmetric solutions. The Bena-Kraus geometry is axially-symmetric and corresponds to a state in the non-twisting sector  of the CFT. It is useful to consider it as the four-dimensional analogue of the two-charge geometry discovered in~\cite{bal,mm}. This two-charge geometry can be regarded as a member of the Lunin-Mathur family and expressed as a particular solution of the chiral-null model~\cite{HorowitzTseytlin}.  This analogy can be taken much further by seeing the two geometries in the context of the proposal of Gaiotto-Strominger-Yin \cite{GSY} relating four-dimensional BPS objects to five-dimensional ones.  The asymptotic behaviour of the two-charge geometry arises from the asymptotic behaviour of the base space -- which in this case is just $\field{R}^4$. On the other hand, Taub-NUT space has an inner region geometry of $\field{R}^4$ and asymptotes to $\field{R}^3 \times S^{1}$. Thus, if one replaces the flat base space of the two-charge geometry with Taub-NUT space one would expect to arrive at the Bena-Kraus solution. The aim of this paper is to show that this construction allows one to use the known D1-D5-P solution in order to add momentum to the D1-D5-KK monopole system of~\cite{bk}. The resulting geometry will be shown to be smooth, up to orbifolding because of the presence of coincident KK monopoles.  
While this work was under way, \cite{BenaWarnernew} and \cite{LeviGimon} appeared, which explored related aspects of the problem.
  
The plan for the rest of the paper is as follows. In \S{2} the ansatz for supersymmetric solutions of minimal supergravity in six dimensions found in~\cite{gmr} is reviewed. Of particular interest are the solutions written on a Gibbons-Hawking base and preserving the isometries of this base. In this case the equations of motion can be explicitly solved in terms of harmonic functions in three dimensions. As a concrete application of the formalism the D1-D5-P solution constructed in~\cite{gms1,gms2} and rewritten in the minimal supergravity ansatz in~\cite{gm1} is reviewed. In \S{3} an ansatz for the six harmonic functions needed to specify the full solution and the regularity conditions to be imposed on the metric and the associated field strength are considered. The resulting constraints can be solved consistently to yield a unique solution with given charges for the KK monopole and the D-branes. In \S{4} the properties of the four charge solution are discussed. The near horizon geometry and asymptotic charges are computed and it is shown that the solution in an appropriate limit reduces to the D1-D5-KK solution found in~\cite{bk}. The paper ends with a discussion of results and open questions.  

\newsection{BPS solutions of minimal supergravity}

Throughout this paper, the framework of $D=6$ minimal supergravity will be used, whose bosonic sector consists of a graviton $g_{\mu\nu}$ and a self-dual three-form. Solutions to this theory can be trivially lifted to Type IIB in ten dimensions by adding a flat $\field{T}^{4}$. This leads to a vanishing dilaton and the three-form may be interpreted as the field strength of the RR 2-form in ten dimensions. All the supersymmetric solutions of this theory were classified by Gutowski, Martelli and Reall in \cite{gmr} by solving the constraints arising from the existence of a Killing spinor in the background. For the current work, it will be only necessary to consider the ansatz describing time independent solutions only:
\bea
ds^2 &=& -2 H^{-1} (du + \beta) \left( dv + \omega + \frac{F}{2} ( du + \beta) \right) + H h_{mn} dx^m dx^n \nonumber \\
G  & =&   *_{4}d H -\frac{1}{2} e^{+} \wedge  ( d\omega - *_{4} d\omega ) + H^{-1} e^{-} \wedge d \beta - H^{-1} e^{+}\wedge e^{-} \wedge d H
\eea
where $ds_{B}^{2} = h_{mn}dx^{m} dx^{n}$ is a metric on a hyperkahler four manifold which will heretofore be referred to as the base space, $H$ and $F$ are functions on the base space while $\omega$ and $\beta$ are one forms on the base space. The hodge dual $*_{4}$ is defined with respect to the metric on the base space. The one-forms $e^{\pm}$ are defined by 
\be
e^{+} = H^{-1} ( du+ \beta ), \ \ e^{-} = dv + \omega + \frac{F}{2} \left( du+ \beta \right)
\ee
For the purpose of this paper, it will be more convenient to trade in the null coordinates $u$ and $v$ for a space-like and time-like coordinate.  Defining
\be
u= \frac{1}{\sqrt{2}  } (t+y),\ v= \frac{1}{\sqrt{2}} (t -y),\ F= 2 (1-Z_{P} ),\  H=Z_{1},\ \beta= \frac{\omega_{P}}{\sqrt{2}},\  \omega = \frac{2k-\omega_{P} }{\sqrt{2}}
\ee
With these redefinitions the metric and the field strength become
\bea
ds^2 &=&  -\frac{1}{Z_{1} Z_{P} }\left( dt + k\right)^2   + \frac{Z_{P}}{Z_{1}} \left( dy +  (1- Z_{P}^{-1} )dt - \frac{k}{Z_{P}}+ \omega_{P}\right)^2 + Z_{1} h_{mn} dx^m dx^n \nonumber \\
G &=&  *_{4} d Z_{1} +  d\left( (dy+ dt+\omega_{P}) \wedge \left[\frac{dt+ k}{Z_{1}} - \omega_{1} \right] \right)  +  \omega_{1}\wedge d\omega_{P}  \label{metans}
\eea
The equations governing the various functions and one forms appearing above are
\bea
d\omega_{1,P} - *_{4} d\omega_{1,P} &=& 0 \nonumber \\
 d *_{4}\! d Z_{1} +   d\omega_{P} \wedge d\omega_{1} &=& 0  \\
 d *_{4}\! d Z_{P}  +  d\omega_{1} \wedge d\omega_{1} &=& 0  \nonumber \\
dk + *_{4} dk  - 2Z_{1} d\omega_{1} - Z_{P} d\omega_{P} &=& 0\nonumber 
\eea
As may be gleaned from the above equations, the possibility of finding explicit solutions depends on the amount of symmetry in the base space. The Gibbons-Hawking (GH) class~\cite{gibbonshawking} provides examples of hyperkahler metrics with an isometry. If the system is further restricted to the case in which this isometry is preserved for the full six-dimensional solution, it was shown in~\cite{bw,gmr} that the above equations can be explicitly solved by dimensional reduction on the isometry direction. The complete solution is then specified by six harmonic functions on three-dimensional flat space. Fortunately the known D1-D5-P solution lies in this class and it will be assumed that this is the case also for the solution under consideration. Start with the Gibbons-Hawking metric in the following form\footnote{$*_{3}$ is performed with respect to the flat three dimensional metric: $ ds_{3}^2 = dr^2 + r^2 d\theta^2 + r^2 \sin^2\theta d\phi^2$ and the orientation on the base space is chosen such that $\epsilon_{z r\theta \phi} >0$ }
\bea
ds_{B}^2&=& h_{mn} dx^m dx^n = V^{-1} (\hat{e})^{2} + V ( dr^2 + r^2 d\theta^2 + r^2 \sin^2\theta d\phi^2 ),\ \hat{e}= dz+ \chi \nonumber \\
d V &=& *_{3} d \chi, \label{ehat}
\eea
The solution is
\bea
\omega_{i}  &=& \vec{\omega}_{i} + \frac{H_{i}}{V} \hat{e},\ d \vec{\omega}_{i} = - *_{3}d H_{i},\ \ d *_{3}\! d H_{i}=0, \ i=1,P \nonumber \\
Z_{1}  &=& \Lambda_{1} + \frac{H_{1} H_{P}}{V},\ \ \  Z_{P} = \Lambda_{P} + \frac{H_{1}^2}{V},\ \ d *_{3}\! d \Lambda_{i}=0,\ i=1,P  \label{gmrsol} \\
k&=&  \vec{k} + k_{0}\hat{e},\ \ k_{0} = H_{k}+  \frac{H_{1} \Lambda_{1}}{V} + \frac{H_{1}^2 H_{P} }{V^2} + \frac{1}{2} \frac{H_{P} \Lambda_{P} }{V} \nonumber,\ d*_{3}\! d H_{k}=0 \\
\ *_{3}d \vec{k} &=& \left( V d  H_{k} - H_{k} d  V \right) +\left( H_{1} d  \Lambda_{1} - \Lambda_{1} d  H_{1} \right) + \frac{1}{2} \left( H_{P} d  \Lambda_{P} - \Lambda_{P} d  H_{P} \right) \nonumber 
\eea
In the case of a two-centre Gibbons-Hawking base which will be the case of interest to us, the two-form gauge potential can be found explicitly.  The resulting expression is
\begin{equation}
B_2 =\left(\tilde{\Lambda}_1-\frac{H_P}{V} \vec{\omega}_1\right) \wedge (dz + \chi) + (dy+dt+\omega_P)\wedge\left(\frac{dt+k}{Z_1}-\omega_1\right)
\end{equation}
where 
\begin{equation}
 d\tilde{\Lambda}_1 \equiv *_3 d\Lambda_1
\end{equation}
The one forms $\chi$, $k$ and $\omega_{i}$ appearing above can also be explicitly found, as shown in Appendix B. 

\subsection{D1-D5-P solution}

A nontrivial application of the above formalism is provided by the D1-D5-P solution constructed in~\cite{gms1,lunin1}. The solution was derived in~\cite{gms1} by taking the extremal limit of a known class of non-extremal solutions~\cite{cy}. These geometries are known to represent bound states since they were obtained by applying exact symmetries (spectral flow and S,T dualities) to 2-charge bound state geometries\footnote{The 2-charge
geometries in turn were known to be bound states because they were generated by S,T dualities applied to a single fundamental string carrying vibrations.}. If one sets $Q_{1}=Q_{5}\equiv Q$ in the three charge solution, they can be embedded in minimal supergravity and therefore one expects them to be of the form of Eq. (\ref{gmrsol}). This was accomplished in~\cite{gm1}. In our notation, the six harmonic functions are set to
\bea
V = \frac{1}{\tgamma_{1} + \tgamma_{2} } \left( \frac{\tgamma_{2} }{r} + \frac{\tgamma_{1} }{r_{c} } \right) \ \ \ \ \ \ \ \ \ \ & & 
H_{1} =  \frac{\tgamma_{1} \tgamma_{2} }{2(\tgamma_{1} + \tgamma_{2} )}  \left( \frac{1}{r_{c} }- \frac{1}{r} \right)  \\
H_{P} = \frac{Q}{2 ( \tgamma_{1} + \tgamma_{2} )} \left( \frac{1}{r }- \frac{1}{r_{c}} \right)\ \ \ \ \ \ \    & & 
\Lambda_{1} =  1+ \frac{Q} {4(\tgamma_{1} + \tgamma_{2} )}\left( \frac{\tgamma_{1} }{r} + \frac{\tgamma_{2} }{r_{c}} \right) \\
\Lambda_{P} = 1- \frac{\tgamma_{1} \tgamma_{2} }{4(\tgamma_{1} + \tgamma_{2} ) } \left( \frac{\tgamma_{1} }{r} + \frac{\tgamma_{2} }{r_{c}}  \right) & &
H_{k} =   \frac{Q}{16( \tgamma_{1} + \tgamma_{2} )} \left( \frac{\tgamma_{1}^2}{r} - \frac{\tgamma_{2}^2 }{r_{c}} \right) 
\eea
where 
\be
r_{c} = \sqrt{r^2 + c^2 + 2 c r \cos\theta}, \ \ c= \frac{ (\tgamma_{1} + \tgamma_{2})^{2} \eta }{4},\ \ \eta= \frac{Q}{ Q - 2 \tgamma_{1} \tgamma_{2} }
\ee
Note the following properties of the above solution which will be useful in the sequel:
\begin{enumerate}
\item From the supergravity point of view, the values of $\tgamma_{1}$ and $\tgamma_{2}$ are arbitrary. However the CFT fixes these values to $-(n+1)Q/R_{y}$ and $nQ/R_{y}$ respectively. This is a consequence of applying spectral flow to the NS ground state. It is remarkable that demanding complete regularity of the supergravity solution also fixes these constants to the same values (except $c$). This fact suggests that in a sufficiently restricted class of solutions (such as the one above with general values of $\tgamma_{1}$ and $\tgamma_{2}$) it might be possible to locate the correct CFT duals without having a detailed knowledge of the CFT. This assumption is important because unlike the D1-D5 system, the D1-D5-KK6 CFT is not known explicitly.
\item The momentum charge $Q_{P}$ is not independent and is set to be $Q_{P} = - \tgamma_{1} \tgamma_{2} $. In view of the previous comment the physical value of the momentum for the relevant states is $n(n+1) Q^2/R_{y}^2$. As before, one can derive this relation from the properties of spectral flow in the $(4,4)$ CFT and it also turns out that this value is required for the supergravity solution to be smooth. In the current work a very similar relation arising for the four charge solution will be found. 
\item In case $n(n+1)=0$ (i.e. $n=0$ or $n=-1$) the momentum vanishes and the solution reduces to the smooth two-charge geometry discovered in~\cite{bal,mm}. It is useful to note that in this case the function $H_{1}$ vanishes and $\Lambda_{P}$ becomes trivial. The base metric reduces to the euclidean metric on $\field{R}^{4}$ (written in GH coordinates with potential $r^{-1}$ or $r_{c}^{-1}$). The analogue of the two-charge geometry for the four-dimensional case is the three-charge metric found by Bena and Kraus in~\cite{bk}. The base metric of this geometry is chosen to be Taub-NUT. There is a close similarity between the five-dimensional two-charge metric and the four-dimensional Bena-Kraus metric when written in terms of the supersymmetric ansatz. This suggests that in order to add the fourth charge to the Bena-Kraus solution one should modify the base metric by adding a second pole to the Gibbons-Hawking base.  
\item The separation $c$ of the two poles in $V$ is related to the charges and the radius of the $y$ circle. It turns out that the singularity analysis cannot fix this constant. The specific value noted above is the one which one arrives at by deconstructing the solution in~\cite{gms1} (which was derived by taking the extremal limit of the non-extremal class). This apparent weakness of the singularity analysis is resolved as follows. The absence of a constant in $V$ allows us to rescale the  coordinates and parameters, in particular the radial coordinate $r\rightarrow \lambda r$  while keeping all the asymptotic charges fixed and sending  $c\rightarrow c/\lambda$. This transformation has the effect of scaling the metric by a constant. This means that the value of $c$ can be freely changed as long as the other parameters are changed simultaneously so as to the keep the physical charges fixed. Hence it is not surprising that one cannot fix this value in supergravity. In the four dimensional case the situation is very different however. In the three charge D1-D5-KK6 solution, for example, this scale invariance is broken by the constant in $V$ and the value of $c$ is uniquely fixed.  
\end{enumerate}

\newsection{Constructing the four-charge geometry}
\subsection{Choice of base space}
The first question with which one is confronted when one tries to generalise the D1-D5-P geometry to include KK charge is the choice of the metric for the base space. One way to address this question is to make use of the recent conjecture of Gaiotto, Strominger and Yin~\cite{GSY} relating four-dimensional and five-dimensional BPS objects. To start, note that Taub-NUT interpolates between $\field{R}^{3}\times S^{1}$ asymptotically and $\field{R}^{4}$ in the core region (up to some discrete identifications). Given a five dimensional object, it may be possible to embed it in the interior of the Taub-NUT geometry by gluing the asymptotic region of the given object to the interior of the Taub-NUT space. If this can be done consistently then the result is an asymptotically four-dimensional object. This idea has been exploited recently to produce a spinning supersymmetric black dihole in four dimensions by embedding the five-dimensional black ring and the BMPV solution in Taub-NUT space~\cite{blackhole4D}. The current work will apply this idea to ``reduce'' the known five-dimensional microstate of the D1-D5-P system to four dimensions and induce the KK-charge. 

The asymptotics of a given solution in the minimal supergravity language depends on the asymptotic behaviour of the function $V$. To get a five-dimensional geometry it is necessary to have $V \sim r^{-1}$ for large $r$ while for a four-dimensional object the leading order behaviour of $V$ needs to be $r^{0}$. Given the connection between 4D and 5D objects, this suggests that the simplest way to embed the D1-D5-P solution in a KK-monopole background is to add a constant to $V$ which will modify the asymptotics of the six dimensional solution to be $\field{R}^{3,1} \times S^{1}\times S^{1}$. Of course, one cannot just add a constant to $V$ without destroying the regularity of the solution and it will be necessary to start with a general ansatz for the remaining five harmonic functions and perform a singularity analysis on the resulting metric and gauge field.

\subsection{The ansatz}
With the previous discussion in mind, one can choose the base metric by specifying the following form for $V$
\be
V= 1+ \frac{Q_{K} }{ \tgamma_{1} +\tgamma_{2} } \left( \frac{\tgamma_{2} }{ r} + \frac{ \tgamma_{1} }{ r_{c}} \right)
\ee
where $r_{c}$ is given by (note that $c$ is arbitrary for now)
\be
r_{c} = \sqrt{r^2+ c^2 + 2 c r \cos\theta}
\ee
In comparison to the function $V$ appearing in the D1-D5-P solution the above choice has a different asymptotic  behaviour because of the presence of $1$. Furthermore the positive constant $Q_{K}$ has been introduced which will later be related to the number of KK monopoles in the solution. Heuristically it is easy to see that this identification is correct by comparing the large distance behaviour of the function to the Taub-NUT space
\be
V\stackrel{\tiny{r\rightarrow \infty}}{\approx} 1+ \frac{Q_{K}}{r}
\ee 
The KK charge $Q_{K}$ is related to the physical number of KK monopoles, $N_{K}$ by
\be
Q_{K} = \frac{1}{2} N_{K} R_{z}
\ee
where $R_{z}$ is the radius of the (compact) fibre direction in the Gibbons-Hawking metric. With this identification the asymptotic behavior of the base metric is $\field{R}^{3} \times S^{1}/\field{Z}_{N_{K}}$. If $\tgamma_{1}$ and $\tgamma_{2}$ were both positive this metric would represent two separate stacks of KK-monopoles at the origin and $r_c=0$ respectively and would have the standard orbifold singularities at those points associated with coincident monopoles. At points away from the poles the metric would be completely smooth. However, the experience with the D1-D5-P system has shown that, as far as black hole microstates are concerned, one must also admit the so-called ``pseudo-hyperkahler'' metrics in which there is a possible signature change from $(4,0)$ to $(0,4)$ at some closed surface in the geometry. This scenario is realised if $\tgamma_{1}$ and $\tgamma_{2}$ have opposite signs. It  will later be found that the same picture is true for the four charge case and therefore $\tgamma_{1}$ and $\tgamma_{2}$ must be allowed to be arbitrary for now. This would make the base metric have severe singularities at the zeroes of $V$. However the full six dimensional geometry can still be smooth if the other harmonic functions are properly chosen. One constraint that suggests itself is that the singularities of the other harmonic functions must coincide with the singularities of the base space. It is not clear whether this is a necessary condition but it is certainly sufficient to allow us to construct a smooth solution\footnote{It is interesting to note that in the zero momentum case both in four and five dimensions the function $V$ has a single pole at the origin whereas the function $\Lambda_{1}$ has a pole only at $r_{c}=0$ while the full six dimensional solution is completely smooth. This would seem to indicate that it is not essential to have all the poles matched between some of the harmonic functions.}. This assumption fixes the singularity structure of the harmonic functions needed to specify the solution. There will be two simple poles at $r=0$ and $r_{c}=0$. Furthermore it is consistent to add an independent constant to each of the functions. However for $H_{1}$ and $H_{P}$ there is a gauge invariance which allows one to fix a single constant in each of the two functions. Consider this in more detail. Recall that $H_{i}$ was related to the one-form $\omega_{i}$ as follows
\be
\ *_{3} d \vec{\omega}_{i} =  - dH_{i}, \ \ i= 1,P
\ee
where
\be
\omega_{i} = \vec{\omega}_{i} + \frac{H_{i} }{ V} (dz+ \chi),\ \ d\chi = *_{3}  d V
\ee
If the replacement $H_{i} \mapsto H_{i} + \alpha_{i} V$ is made, where $\alpha_{i}$ is an arbitrary real number, one finds
\be
\vec{\omega}_{i} \mapsto \vec{\omega}_{i} - \alpha_{i} \chi
\ee
and hence the complete one form $\omega_{i}$ changes to
\be
\omega_{i} \mapsto \omega_{i} + \alpha_{i} dz
\ee
If $i=P$ it is seen from the metric and the field strength Eq. (\ref{metans}) that this change in $\omega_{P}$ can be removed by a redefinition of $y$. On the other hand $\omega_{1}$ appears only in the field strength $G$ through its own field strength $d\omega_{1}$ which is invariant under the above change. This demonstrates the invariance of the solution under the above gauge transformation. In writing down the ansatz one can make use of this invariance to simplify the solution. It will be convenient to choose $\alpha_{1}$ so that $H_{1}$ has no constant term. For $H_{P}$ it will be best to  make the choice so as to set the sum of the residues of the two poles at $r=0$ and $r_{c}=0$ to zero. 

The above considerations lead to the following ansatz
\bea
V =1+ \frac{Q_{K}}{\tgamma_{1} + \tgamma_{2} } \left( \frac{\tgamma_{2} }{r} + \frac{\tgamma_{1} }{r_{c} } \right)  \, \ \ \ \ \ & & 
H_{1} = \frac{\tgamma_{1} \tgamma_{2}}{\tgamma_{1}+\tgamma_{2} } \left( \frac{b_{12}}{r} + \frac{b_{22}}{r_{c}} \right)\\
H_{P} = c_{1} + \frac{c_{12}}{r} + \frac{c_{22}}{r_{c}} \ \ \ \ \ \ \ \ \ \ \ \ \ \ \ \ \    & & 
\Lambda_{1} =  1+ \frac{ \lambda_{12} }{r} + \frac{\lambda_{22}}{r_{c}}  \\
\Lambda_{P} = 1+ \frac{ \mu_{1} }{r} + \frac{\mu_{2}}{r_{c}} \ \ \ \ \ \ \ \ \ \ \ \ \ \ \ \ \ \ & &
H_{k} =   h_{k} + \frac{ q_{k1}}{r} + \frac{q_{k2}}{r_{c}} 
\eea
As a consequence of the gauge invariance $c_{12}=-c_{22}$. However, it is better to leave this condition implicit for now because the equations arising from the singularity analysis acquire a more symmetric form. Note also that the constant $c$ has been left undetermined. Later, it will be found to be related to the radius of the $y$-circle. 

In the next section, the regularity conditions which need to be imposed in order to have a smooth solution are considered. These conditions will fix the arbitrary constants $b_{i2},c_{i2},\lambda_{i2},\mu_{i},q_{ki},\ i=1,2$ and $h_{k}$ in the ansatz above in terms of the asymptotic charges and the radius of the $y$-circle. 

\subsection{The singularity analysis}
Consider the following conditions that need to be imposed:
\begin{enumerate}
\item The determinant of the metric should vanish at $r=0$ and $r_c=0$. The (square root of the) determinant is given by
\be
\sqrt{|g|} = Z_{1}  V r^2 \sin\theta
\ee
The condition above then implies that $Z_{1} V$ should not have terms of order $r^{-2}$. 
\be
\lambda_{12} = - \frac{\tgamma_{1}b_{12} c_{12}}{Q_{K}} , \ \ \lambda_{22} = - \frac{\tgamma_{2} b_{22}  c_{22}}{Q_{K}} \label{lambda1222}
\ee
For $Z_{P}$ one has to ensure that there are no singularities at either $r=0$ or $r_{c}=0$, otherwise the $dt^2$ and $dy^2$ components would be singular.  This yields 
\be
\mu_{1} = -\frac{\tgamma_{1}^2 \tgamma_{2} b_{12}^2}{Q_{K}(\tgamma_{1} + \tgamma_{2})} ,\ \  \mu_{2} = - \frac{\tgamma_{1}\tgamma_{2}^2 b_{22}^2}{Q_{K} (\tgamma_{1} + \tgamma_{2} )}\label{mu1mu2}
\ee
These conditions then completely determine $\Lambda_{1}$ and $\Lambda_{P}$ in terms of the constants $b_{i2},c_{i2}$:
\bea
\Lambda_{1} &=&  1-\frac{ \tgamma_{1} b_{12} c_{12} }{Q_{K} r} - \frac{\tgamma_{2}b_{22} c_{22} }{ Q_{K} r_{c} }  \label{fullL1} \\
\Lambda_{P} &=&  1 - \frac{\tgamma_{1} \tgamma_{2}}{Q_{K} (\tgamma_{1} + \tgamma_{2} )} \left( \frac{\tgamma_{1}  b_{12}^2}{r}  + \frac{\tgamma_{2} b_{22}^2}{r_{c}}\right) \label{fullLp}
\eea
\item $\vec{k}$ should vanish at $\theta=0,\pi$ and all values of $r$. This condition ensures that $\theta=0$ and $\theta=\pi$ are the two poles of a two sphere. $\vec{k}$ can be calculated as in Appendix B:
\be
\vec{k} = \left\{ f_{1} \cos\theta  + f_{2}  \frac{r \cos\theta + c }{r_{c} }  + f_{3} \frac{ r_{c} -r  - c \cos\theta }{ c r_{c} } \right\}  d\phi \label{defnk}
\ee
where
\bea
f_{1} &=& q_{k1} -\frac{1}{2} c_{12} - \frac{Q_{K} \tgamma_{2}  }{\tgamma_{1}+ \tgamma_{2}} h_{k} - \frac{\tgamma_{1}\tgamma_{2} }{\tgamma_{1}+ \tgamma_{2}} \left( b_{12} + \frac{\tgamma_{1} c_{1} b_{12}^2 }{2 Q_{K}} \right)  \label{f1}  \\
f_{2} &=&  q_{k2} -\frac{1}{2} c_{22} - \frac{Q_{K} \tgamma_{1}  }{\tgamma_{1}+ \tgamma_{2}} h_{k} - \frac{\tgamma_{1}\tgamma_{2} }{\tgamma_{1}+ \tgamma_{2}} \left( b_{22} + \frac{\tgamma_{2} c_{1} b_{22}^2 }{2 Q_{K}} \right) \label{f2} \\
f_{3} &=& \frac{Q_{K} (\tgamma_{2} q_{k2} - \tgamma_{1} q_{k1} ) }{\tgamma_{1}+\tgamma_{2}} \nonumber \\ &&  + \frac{\tgamma_{1} \tgamma_{2}}{Q_{K}(\tgamma_{1}+\tgamma_{2})} \left[ b_{12} b_{22} \left( \tgamma_{1} c_{12} - \tgamma_{2}c_{22}  \right) + \frac{1}{2} \left(  \tgamma_{1} b_{12}^2 c_{22} - \tgamma_{2} b_{22}^2 c_{12}  \right)\right]\label{f3}
\eea
 The condition at $\theta=0$ gives for all values of $r$
 \be
 f_{1} + f_{2} =0 \label{eqnone}
 \ee
 The condition at $\theta=\pi$ and $r>c$ yields the same equation as above while the one for $r<c$ yields
 \be
 f_{1} - f_{2} -  \frac{2f_{3}}{c}=0 \label{eqntwo}
 \ee
 \item $\vec{k}$ and $ k_{0}$ should vanish at $r=0$. This condition is necessary to ensure that $r=0$ is a point and there is no singularity there. Around $r=0$, $\vec{k}$ is finite and gives the condition
 \be
\left( f_{1} - \frac{f_{3}}{c} \right)  \cos\theta d\phi + \left( f_{2} +\frac{f_{3}}{c} \right) d\phi =0
\ee
which is the same condition as the two previous ones. In this limit, $k_{0}$ has a term at order $r^{-1}$ and a constant piece. The divergent piece yields the following simple condition
\be
q_{k1} = \frac{\tgamma_{1}^2 b_{12}^2 c_{12}}{2 Q_{K}^2} \label{qk1}
\ee
while the constant piece leads to
\bea
\!\!\!\!\!\!\!\!\!\!\!\!\!\! & & h_{k} + \frac{q_{k2}}{c} + \frac{\tgamma_{1} b_{12}}{Q_{K}} \left( 1 - \frac{\tgamma_{2} b_{22 } c_{22} }{cQ_{K} } \right) + \frac{\tgamma_{1}^2 b_{12} b_{22} c_{12}}{c Q_{K}^2}  + \frac{  (\tgamma_{1} + \tgamma_{2}) c_{12} }{2 Q_{K}  \tgamma_{2} } \left( 1 - \frac{\tgamma_{1} \tgamma_{2}^2 b_{22}^2}{ c Q_{K} (\tgamma_{1}+\tgamma_{2}) } \right) \nonumber  \\ \!\!\!\!\!\!\!\!\!\!\!\!\!\! & & 
 + \frac{ \tgamma_{1}^2 b_{12}^2 }{2Q_{K}^2 } \left( c_{1} + \frac{c_{22} }{c} \right)   - \frac{\tgamma_{1}^2(\tgamma_{1}+\tgamma_{2}) b_{12}^2 c_{12} }{2\tgamma_{2} Q_{K}^3} \left( 1+ \frac{Q_{K} \tgamma_{1}}{c(\tgamma_{1}+\tgamma_{2})} \right) =0  \label{eqnthree}
\eea
\item $\vec{k}$ should be finite as $r_{c} \rightarrow 0$ and $k_{0}$ should vanish in the same limit. This is necessary because as $r_{c}\rightarrow 0$ the $z$ circle shrinks to zero size. The finiteness of $\vec{k}$ does not yield any new conditions but vanishing of $k_{0}$ gives
\be
q_{k2} = \frac{\tgamma_{2}^2 b_{22}^2 c_{22}}{2 Q_{K}^2}  \label{qk2}
\ee
\bea
\!\!\!\!\!\!\!\!\!\!\!\!\!\! & & h_{k} + \frac{q_{k1}}{c} + \frac{\tgamma_{2} b_{22}}{Q_{K}} \left( 1 - \frac{\tgamma_{1} b_{12 } c_{12} }{cQ_{K} } \right) + \frac{\tgamma_{2}^2 b_{12} b_{22} c_{22}}{c Q_{K}^2}  + \frac{  (\tgamma_{1} + \tgamma_{2}) c_{22} }{2 Q_{K}  \tgamma_{1} } \left( 1 - \frac{\tgamma_{1}^2 \tgamma_{2}  b_{12}^2}{ c Q_{K} (\tgamma_{1}+\tgamma_{2}) } \right) \nonumber  \\ \!\!\!\!\!\!\!\!\!\!\!\!\!\! & & 
 + \frac{ \tgamma_{2}^2 b_{22}^2 }{2Q_{K}^2 } \left( c_{1} + \frac{c_{12} }{c} \right)   - \frac{\tgamma_{2}^2(\tgamma_{1}+\tgamma_{2}) b_{22}^2 c_{22} }{2\tgamma_{1} Q_{K}^3} \left( 1+ \frac{Q_{K} \tgamma_{2}}{c(\tgamma_{1}+\tgamma_{2})} \right) =0  \label{eqnfour}
\eea

\item Behavior of $\vec{\omega}_{P}$ at $\theta=0,\pi$: $\vec{\omega}_{P}$ is given by
 \be
 \vec{\omega}_{P} = - c_{12} \cos\theta d\phi - c_{22} \frac{r \cos\theta + c}{r_{c} } d\phi  \label{omegavecP}
 \ee
At $\theta=0,\pi$ the $\phi$ cycle shrinks to zero size and for regularity the condition that $\vec{\omega}_{P}$ is zero might be imposed. However due to the mixing with $y$ cycle the following weaker condition may be imposed and still result in regularity. Demand that as $\theta\rightarrow 0,\pi$ the vector field $\vec{\omega}_{P}$ should be a half-integer multiple of the radius of the $y$ circle: 
\be
c_{12} = \frac{1}{2}m_{1} R_{y},\ c_{22} = \frac{1}{2}m_{2} R_{y},\ \ m_{1},m_{2} \in \field{Z} 
\ee
As discussed a little earlier, by making use of the gauge freedom in the definition of $H_{P}$ the constraint $m_{1}+ m_{2}=0$ can be set, which simplifies the computations. Thus it is useful to define
\be
c_{22} = - c_{12}= -\frac{1}{2} m R_{y},\ \  m \in \field{Z}_{+} \label{c12c22}
\ee
where $R_{y}$ is the radius of the $y$-circle. 

\item Regularity of the base space: In the D1-D5-P case the base metric had severe singularities at $r=0$ and $r_{c}=0$ in general. However for special values of $\tgamma_{1}$ and $\tgamma_{2}$ which were determined by the CFT, $r,r_{c}=0$ were mild orbifold singularities which were resolved in the full geometry. The base metric for the four charge system is closely related to the three-charge case and similar constraints must be imposed on $\tgamma_{1}$ and $\tgamma_{2}$ so as to admit only orbifold singularities which will be resolved in the full metric. From the structure of $\chi$ in Eq. (\ref{chieqn}) it is clear that one should impose
\be
\frac{\tgamma_{i} }{\tgamma_{1} + \tgamma_{2} } \equiv n_{i} \in \field{Z} , \ \ i=1,2
\ee
which is easily solved to yield
\be
\tgamma_{1} = -n  , \ \tgamma_{2} =  n+1 , \ \ n\in \field{Z} \label{valueofgamma}
\ee 

The orbifold singularity of the base is resolved in the full space. To see this
consider the behaviour of $dy+\omega_P$ around one of the poles, say $r=0$:
\bea
dy+\omega_P&\approx& dy +c_{12}\,{\tgamma_1+\tgamma_2\over \tgamma_2}\,{dz\over Q_K}+\Bigl(c_{12}\,{\tgamma_1\over \tgamma_2}-c_{22}\Bigl)\,d\phi\nonumber\\
&=&dy+{ m R_y\over 2(n+1)}\,\Bigl({dz\over Q_K}+d\phi\Bigr)
\label{ymixing}
\eea 
where in the last line the previously found values for $c_{12}$, $c_{22}$, $\tgamma_1$ and $\tgamma_2$ have been used. The orbifold singularity of the base space at $r=0$ originates from the fact that the coordinates
$dz/Q_K$ and $d\phi$ shrinks to a point, giving rise to a fixed point of the 
orbifold group $\mathbb{Z}_{n+1}$. The equation (\ref{ymixing}) shows that 
$dy$ mixes with $dz/Q_K$ and $d\phi$ precisely in such a way as to resolve 
the orbifold point.

\item Regularity of the gauge field: It is convenient to rewrite
the RR field strength as
\be
G =  *_{4} d Z_{1} +  d\left( (dy+ dt+\omega_{P}) \wedge
 \frac{(dt+ k)}{Z_{1}}\right)  + (dy+ \omega_{P})\wedge  d \omega_{1}
\ee 
The conditions imposed above guarantee that $Z_1$, $dy+\omega_P$ and
$dt+k$ are everywhere well-defined. Thus it is only necessary to make sure that 
$d\omega_1$ be well-defined. At $\theta=0,\pi$ it should be true that 
$d\vec{\omega}_1=0$. But 
\be
d\vec{\omega}_1={\tgamma_1\tgamma_2\over \tgamma_1+\tgamma_2}\Bigl(b_{12}\,
\sin\theta\,d\theta+b_{22}\,
{c\,r\,\sin^2\theta\over r_c^3}\,dr+b_{22}\,
{r^2\,(r+c\cos\theta)\,\sin\theta\over r_c^3}\,d\theta\Bigr)\,d\phi
\ee 
and thus $d\vec{\omega}_1$ vanishes at $\theta=0,\pi$ for any value of $b_{12}$
and $b_{22}$. By direct inspection one can also check that $d\omega_1$ is
well-defined at $r=0$ and $r_c=0$ for any $b_{12}$ and $b_{22}$. One further
condition comes by demanding that the field strength vanishes at asymptotic 
infinity: this requires that $\vec{\omega}_1\rightarrow 0$ as 
$r\rightarrow \infty$ which leads to
\be
b_{22} = - b_{12} \equiv b_{2} \label{b12b22}
\ee

\item $k_{0}\rightarrow 0$ as $r\rightarrow \infty$: This condition should be imposed in order to acheive asymptotic flatness in four dimensions. This leads to
\be
 h_{k} + \frac{1}{2}  c_{1} = 0 \label{eqnfive}
\ee
\end{enumerate}

\subsection{Solving the constraints}
The regularity conditions imposed above are sufficient to determine all the unknown parameters. Some of the coefficients are easily determined. Using Eqs. (\ref{lambda1222},\ref{mu1mu2}) $\Lambda_{1}$ and $\Lambda_{P}$ can be completely fixed as in (\ref{fullL1},\ref{fullLp}). Furthermore from Eqs. (\ref{qk1}), (\ref{qk2}), (\ref{c12c22}), (\ref{b12b22})  and Eq. (\ref{eqnfive}) one finds
\be
q_{k1} =  -\frac{\tgamma_{1}^2 b_{2}^2 c_{22}}{2 Q_{K}^2}, \ q_{k2}= \frac{\tgamma_{2}^2 b_{2}^2 c_{22}}{2 Q_{K}^2}, \ c_{1} = - 2 h_{k} 
\ee
Thus the system is left with unknowns $h_k, b_{2}$ and the separation between the two poles $c$. It will be found that $c$ is related to $R_{y}$ and hence to $c_{22}$.  The nontrivial equations which constrain these variables are Eq.(\ref{eqnone}),(\ref{eqntwo}),(\ref{eqnthree}) and (\ref{eqnfour}).  It would seem that the system is over-constrained but it turns out that there are only two independent equations. They can be used to determine $h_{k}$ and $c$ in terms of the free parameters of the solution. The remaining parameter $b_{2}$ which seems to stay unfixed corresponds to an independent D1-brane charge $Q_{1}$ (which determines the D5 charge because $Q_{5}=Q_{1}$). Therefore, eliminate $b_{2}$ in favor of $Q_{1}$. The asymptotic behavior of the function $Z_{1}$ is
\be
Z_{1} \approx 1  - \frac{(\tgamma_{1} + \tgamma_{2} ) b_{2}  c_{22} }{Q_{K} r} + \mathcal{O}(r^{-2} ) 
\ee
which motivates the definition
\be
Q_{1}  \equiv  - \frac{(\tgamma_{1} + \tgamma_{2} ) b_{2}  c_{22} }{Q_{K} }
\ee
In the next section it will be shown that $Q_{1}$ is proportional to the physical D1-brane charge. For now, use the above equation to express $b_{2}$ in terms of $Q_{1}$ and the radius of the $y$-circle $R_{y}$. 
\be
 b_{2} =  \frac{ 2 Q_{1} Q_{K} }{ m(\tgamma_{1} + \tgamma_{2} )  R_{y} }
\ee
The above result will be used to to eliminate $b_{2}$ from the remaining equations. Solving Eq. (\ref{eqnone}) for $h_{k}$,
\be
h_{k}= -\frac{ m (\tgamma_{2}^2 - \tgamma_{1}^2 ) Q_{1}^2 R_{y} }{m^2 (\tgamma_{1} + \tgamma_{2})^2 Q_{K} R_{y}^2 - 4 \tgamma_{1} \tgamma_{2} Q_{1}^2 Q_{K} }
\ee
Substituting this result in the remaining three equations, one finds that they are dependent. Specifically $\tgamma_{2}$ times Eq. (\ref{eqnthree}) added to $\tgamma_{1}$ times Eq. (\ref{eqnfour}) is zero. Furthermore, the other linear combination of these two equations is equal to the negative of Eq. (\ref{eqntwo}). Thus one can choose to solve any one of the three equations; for concreteness, choose to work with Eq. (\ref{eqntwo}). The equation is linear in $c$ and fixes the separation between the two poles in terms of the free parameters of the solution
\be
c =  \frac{ 4 (\tgamma_{1} + \tgamma_{2})^2  \left[ ( \tgamma_{1} + \tgamma_{2})^2 m^2 R_{y}^2 - 4 \tgamma_{1} \tgamma_{2} Q_{1}^2 \right] Q_{1}^2 Q_{K} }{16 \tgamma_{1}^2 \tgamma_{2}^2 Q_{1}^3 ( Q_{1} + 2 Q_{K} ) - 4 (\tgamma_{1} + \tgamma_{2})^2 m^2 Q_{1} \left[  ( \tgamma_{1}^2 + \tgamma_{2}^2) Q_{1} + 2 \tgamma_{1} \tgamma_{2} Q_{K} \right] R_{y}^2 + ( \tgamma_{1} +\tgamma_{2})^4 m^4 R_{y}^4 }
\ee
At this point all the arbitrary constants are fixed and one has the complete solution. It will be illuminating to summarize the solution that has been obtained:
\bea
V &=& 1+ Q_{K} \left( \frac{ n+1 }{r} - \frac{n }{r_{c} } \right) \label{beginsoln} \\
H_{k} &=& -\frac{ m (2n+1) Q_{1}^2 R_{y} }{m^2  Q_{K} R_{y}^2 + 4 n(n+1) Q_{1}^2 Q_{K} }  + \frac{ Q_{1}^2}{ m R_{y} } \left( \frac{n^2 }{ r} - \frac{(n+1)^2 }{r_{c}} \right)    \\ 
\Lambda_{1} &=& 1 +  Q_{1} \left( \frac{n+1}{r_{c}} -\frac{n }{ r}  \right)\\ 
H_{1} &=&  \frac{ 2n(n+1) Q_{1} Q_{K}  }{m R_{y} }\left( \frac{1}{r} - \frac{1}{r_{c}} \right)\\
\Lambda_{P} &=& 1+ \frac{4n(n+1)  Q_{1}^2 Q_K }{ m^2 R_{y}^2 } \left( \frac{n+1}{r_{c}} -\frac{n }{r}  \right)\\
H_{P} &=&   \frac{ 2 m (2n+1 ) Q_{1}^2 R_{y} }{m^2 Q_{K} R_{y}^2 + 4 n (n+1) Q_{1}^2 Q_{K} }  + \frac{m R_{y}}{2} \left( \frac{1}{r } - \frac{1}{r_{c}} \right) \\
k_{0} &=& H_{k} + \frac{\Lambda_{1} H_{1} }{V} + \frac{\Lambda_{P} H_{P}}{2 V} + \frac{H_{1}^2 H_{P} }{V^2}  \\
k &=& - \frac{Q_{1}^2 Q_{K} }{m c R_{y} } \left( 1- \frac{r+c}{r_{c} } \right) \left( 1+ \cos\theta \right) d\phi + k_{0} \hat{e} \\
\omega_{P} &=& \frac{1}{2} m R_{y} \left[ \frac{ (r - r_{c} ) \cos\theta  + c }{ r_{c} } \right] d\phi + \frac{H_{P}}{V} \hat{e} \\
\omega_{1} &=& -\frac{2 n(n+1) Q_{1} Q_{K} }{ m  R_{y} } \left[ \frac{ (r_{c} - r ) \cos\theta  + c }{ r_{c} } \right] d\phi + \frac{H_{1}}{V} \hat{e} \\
\hat{e} &\equiv & dz + Q_{K} \left( (n+1) \cos\theta -n \frac{ r \cos\theta + c }{ r_{c}} \right) d\phi 
\eea
with
\be
c =  \frac{ 4 Q_{1}^2 Q_{K}\left[ m^2 R_{y}^2 + 4 n(n+1) Q_{1}^2 \right]  }{16 n^2 (n+1)^2Q_{1}^3 ( Q_{1} + 2 Q_{K} ) - 4  m^2 Q_{1} \left[Q_{1}+  2n(n+1) (Q_{1}-Q_{K})  \right] R_{y}^2 + m^4 R_{y}^4 } \label{endsoln}
\ee

\subsection{Smoothness of the metric}

Having derived the solution above, it is instructive to see the regularity of the complete metric. There are potential singularities  at $r,r_{c}=0$ and $V=0$. Throughout this section it will be assumed that $\tgamma_1 \tgamma_2 \neq 0$. The remaining case in which either $\tgamma_{1}$ or $\tgamma_{2}$ is zero is the three charge D1-D5-KK6 metric. It will demonstrated in the next section that the current solution reduces in this limit to the Bena-Kraus solution and hence is smooth as well.  

\subsubsection{$r\rightarrow 0$}
Start by considering the behavior of the base metric around $r=0$: 
\be
ds^2_{B} \approx \frac{r \left( dz + Q_{K} \left( (n+1) \cos\theta - n  \right) d\phi \right)^2}{(n+1)Q_{K}} +  (n+1)Q_{K}  \left( \frac{dr^2}{r} + r d\theta^2 + r \sin^2\theta d\phi^2 \right)
\ee
If one defines
\be
\rho^2 = 4 (n+1)Q_{K}  r , \ \bar{\theta} = \frac{\theta}{2}, \ \tilde{\psi} =  \frac{z + Q_{K} \phi }{2(n+1)Q_{K}  },\ \tilde{\phi} =  \frac{(2n+1) Q_{K} \phi - z }{2(n+1) Q_{K}} \label{r0desing}
\ee
the base metric looks like the euclidean metric on $\field{R}^4$ in the new coordinates
\be
ds^2_{B} \approx d\rho^2 + \rho^2 \left( d\bar{\theta}^2 + \cos^2\bar{\theta} d\tilde{\psi}^2 + \sin^2\bar{\theta} d\tilde{\phi}^2 \right)
\ee
With the current choice of gauge for the KK 1-form $\chi$ the coordinates $z$ and $\phi$ are subject to the identifications
\be
(z, \phi) \cong (z, \phi)+2\pi\,n_1\,(R_z, 0)+2\pi\,n_2\Bigl({R_z\,N_K\over 2}
, 1\Bigr)
\ee
with the radius of the $z$ circle given by
\be
R_z=2\,{Q_K\over N_K}
\ee
These induce the following identifications on $\tilde{\psi}$ and $\tilde{\phi}$
\bea
(\tilde{\psi}, \tilde{\phi})&\cong& (\tilde{\psi}, \tilde{\phi})+ {2\pi\,n_1\over N_K}\,\Bigl({1\over n+1},-{1\over n+1}\Bigr)+2\pi\,n_2\,\Bigl({1\over n+1},1-{1\over n+1}\Bigr)\nonumber\\
&\cong& (\tilde{\psi}, \tilde{\phi})+ {2\pi\,n_1\over N_K}\,\Bigl({1\over n+1},-{1\over n+1}\Bigr)+2\pi\,n'_2\,(0,1)
\label{r0orb}
\eea
The base space around $r=0$ is thus a $\mathbb{Z}_{N_K (n+1)}$ orbifold of
flat space: the orbifold is singular because $r=0$, where both $\tilde{\psi}$ and $\tilde{\phi}$ shrink to zero size, is a fixed point of the 
$\mathbb{Z}_{N_K (n+1)}$ action defined by (\ref{r0orb}).

Next, consider the full six dimensional geometry in the same limit. The functions $Z_{1}$ and $Z_{P}$ are regular and approach constants which will be called $\alpha_{1}$ and $\alpha_{P}$ respectively\footnote{For $n>0$, the signature of the base metric around $r=0$ is $(4,0)$. For the full six-dimensional geometry to have signature $(5,1)$ it is necessary that
$\alpha_1$ be positive. Using the explicit form of the solution given above
one obtains the following expression for $\alpha_1$
\bea
&&\alpha_1=(4 (n+1) Q_1 Q_K\,[m^2 R_y^2 +4 n(n+1) Q_1^2])^{-1}
\Bigl[(m^2 R_y^2 -2(n+1) Q_1^2)^2 + 4 Q_1^4(n+1)^2(4n(n+1)-1)\nonumber\\
&&\qquad +4 (n+1)(2n+1)
m^2 R_y^2 Q_1 Q_K +16 n (n+1)^2 (2n+1) Q_1^3 Q_K\Bigr]
\eea
which shows that $\alpha_1$ is positive for any $n\ge 1$.}. Furthermore the regularity conditions imposed earlier guarantee that the one form $k$ is zero in this limit. Thus  
\be
ds^2 \approx - \frac{1}{\alpha_{1} \alpha_{P}} dt^2 + \frac{\alpha_{P} }{\alpha_{1} } \left[ dy + (1- \alpha_{P}^{-1} ) dt +{ m R_y\over 2(n+1)}\,\Bigl({dz\over Q_K}+d\phi\Bigr) \right]^{2} + \alpha_{1} ds^2_{B}
\ee
Defining
\be
\tilde{y} = y +{m R_y\over 2(n+1)}\,\Bigl({z\over Q_K}+ \phi\Bigr) 
\ee
and switching from $(y,z,\phi)$ coordinates to $(\tilde{y},\tilde{\psi},\tilde{\phi})$ coordinates the metric becomes
\be
ds^2 \approx - \frac{1}{\alpha_{1} } (dt^2- d\tilde{y}^2) - \frac{1- \alpha_P }{\alpha_{1}} \left( dt + d\tilde{y}\right)^2 + \alpha_{1}\left(d\rho^2 + \rho^2 d\bar{\theta}^2 + \rho^2\cos^2\bar{\theta} d\tilde{\psi}^2 + \rho^2\sin^2\bar{\theta} d\tilde{\phi}^2 \right)
\ee  
The new coordinates have the identifications
\bea
\Bigl({\tilde{y}\over R_y},\tilde{\psi},\tilde{\phi}\Bigr) &\cong & 
\Bigl({\tilde{y}\over R_y},\tilde{\psi},\tilde{\phi}\Bigr) + 
{2 \pi \,n_1\over N_K}\,
\Bigl({m\over n+1},{1\over n+1},-{1\over n+1}\Bigr)\\
&&\qquad \qquad\quad +2\pi\,n_2\,
\Bigl({m\over n+1},{1\over n+1},1-{1\over n+1}\Bigr)+2\pi\,n_3(1,0,0)\nonumber\\&\cong & \Bigl({\tilde{y}\over R_y},\tilde{\psi},\tilde{\phi}\Bigr) + 
{2 \pi \,n_1\over N_K}\,
\Bigl({m\over n+1},{1\over n+1},-{1\over n+1}\Bigr)\nonumber\\
&&\qquad \qquad\quad +2\pi\,n'_2\,
(0,0,1)+2\pi\,n_3(1,0,0)\nonumber
\eea
As the $\tilde y$ circle remains of finite size at $r=0$, the orbifold fixed point is resolved if the integer $m$ does not share a common factor with
$N_k (n+1)$. If $m$ has a common factor $l$ with $N_k (n+1)$, the subgroup 
$\mathbb{Z}_l$ of $\mathbb{Z}_{N_K (n+1)}$ has a fixed point at $r=0$ and 
the metric displays an orbifold singularity of order $l$.

\subsubsection{$r_{c} \rightarrow 0$}
Now, turn to the limit of $r_{c} \rightarrow 0$. It is convenient to change coordinates so that $r=c,\theta= \pi$ becomes the origin. This is accomplished by the following change of coordinates
\be
r'\cos\theta' = r \cos\theta + c, \ \  r'\sin\theta' = r\sin\theta \label{rcorigin}
\ee
We note a few algebraic identities which are useful in executing this coordinate change
\be
r= \sqrt{ r'^2 + c^2 -2 c r' \cos\theta' } \equiv r'_{c}, \ r_{c} = r', \ \ \frac{r \cos\theta + c}{r_{c}} = \cos\theta'
\ee 
The base metric near $r' =0$  is approximately
\be
ds^2_{B}\approx  -\frac{r' }{n Q_{K} } \left( dz- Q_{K} ( n+1 + n \cos\theta' )  d\phi \right) ^2   - \frac{n Q_{K}}{r' }  \left( dr'^2 + r'^2 d\theta'^2 + r'^2 \sin^2\theta' d\phi^2 \right)
\ee
Note that, assuming $n>0$ for definiteness,  the signature of the metric has flipped from $(4,0)$ to $(0,4)$. This is the same behavior as observed in the D1-D5-P case in~\cite{gm1}. A change of coordinates similar to Eq. (\ref{r0desing}),
\be
\rho'^2 = 4 n Q_{K}  r' , \ \bar{\theta}' = \frac{\theta'}{2}, \ \tilde{\psi}' =  \frac{z - (2n+1) Q_{K} \phi }{2nQ_{K}  },\ \tilde{\phi}' =  \frac{z- Q_{K} \phi }{2 n  Q_{K}} \label{rcdesing}
\ee
diagonalizes the above metric:
\be
ds^2_{B} \approx -d\rho'^2 - \rho'^2 \left( d\bar{\theta}'^2 + \cos^2\bar{\theta}' d\tilde{\psi}'^2 + \sin^2\bar{\theta}' d\tilde{\phi}'^2 \right)
\ee
Even though the base metric has switched signatures the full six dimensional geometry preserves its signature. This is possible because the function $Z_{1}$ 
 approaches a negative constant as $r' \rightarrow 0$. Denote this 
constant value by $-\alpha'_{1}$ where $\alpha'_{1}>0$\footnote{This follows
from the explicit form of $-\alpha'_{1}$ (which is related to the one for 
$\alpha_1$ by the exchange $n\to -(n+1)$):
\bea
&&-\alpha'_1=(4 n Q_1 Q_K\,[m^2 R_y^2 +4 n(n+1) Q_1^2])^{-1}
\Bigl[(m^2 R_y^2 +2 n Q_1^2)^2 + 4 Q_1^4 n^2(4n(n+1)-1)\nonumber\\
&&\qquad +4 n (2n+1)
m^2 R_y^2 Q_1 Q_K +16 n^2 (n+1) (2n+1) Q_1^3 Q_K\Bigr]
\eea
 }. 
Also, 
denote by $-\alpha'_{P}$ the limit of $Z_P$ for $r' \rightarrow 0$.
Then the full geometry is
\be
ds^2 \approx - \frac{1}{\alpha'_{1} \alpha'_{P}} dt^2 + \frac{\alpha'_{P} }{\alpha'_{1} } \left[ dy + (1+ \left.\alpha'\right._{P}^{-1} ) dt +{ m R_y\over 2n}\,\Bigl({dz\over Q_K}- d\phi\Bigr) \right]^{2} + \alpha'_{1} ( d\rho'^2 + \rho'^2 d\Omega_{3}^2 )
\ee
Defining as before
\be
\tilde{y}' = y +{m R_y\over 2n}\,\Bigl({z\over Q_K} - \phi\Bigr) 
\ee
and rearranging the metric in the coordinates $( \tilde{y}', \tilde{\psi}', \tilde{\phi}' )$
\be
ds^2 \approx + \frac{1+ \alpha'_P }{\alpha'_{1}} \left( dt + d\tilde{y}'\right)^2 + \frac{1}{\alpha'_{1} } (dt^2- d\tilde{y}'^2) + \alpha'_{1}\left(d\rho'^2 + \rho'^2  d\bar{\theta'}^2 + \rho'^2\cos^2\bar{\theta}' d\tilde{\psi}'^2 + \rho'^2\sin^2\bar{\theta}' d\tilde{\phi}'^2 \right)
\ee  
with coordinate identifications
\bea
\Bigl({\tilde{y}'\over R_y},\tilde{\psi}',\tilde{\phi}'\Bigr) 
\cong \Bigl({\tilde{y}'\over R_y},\tilde{\psi}',\tilde{\phi}'\Bigr) + 
{2 \pi\,n_1\over N_K}\,\Bigl({m\over n},{1\over n},{1\over n}\Bigr)+
2\pi\,n_2\,(0,1,0)+2\pi\,n_3\,(0,0,1) 
\eea
According to the same logic as in the $r\to 0$ case, it is clear that the full
geometry is regular at $r_c=0$ if $m$ does not have common factors with $nN_K$. It has an orbifold singularity of order $l'$ if $m$ and $nN_K$ share the factor $l'$.

\subsubsection{$V\rightarrow 0$}
Another place to look for singularities is anywhere the function $V$ vanishes.  Note that for generic values of $\tilde{\gamma}_1$ and $\tilde{\gamma}_2$, the six harmonic functions are independent, so none of the other functions vanish at the same points as $V$.  To investigate such a possibility, write the functions and one-forms appearing in the solution as expansions in $1/V$:
\begin{equation}
Z_1 = \frac{H_1 H_P }{V}+ \Lambda_1,
Z_P = \frac{H_1^2}{V}+\Lambda_P
\end{equation}
\begin{equation}
\omega_P = \frac{H_P}{V}\hat{e} +\vec{\omega}_P,
\omega_1 = \frac{H_1}{V}\hat{e} +\vec{\omega}_1
\end{equation}
\begin{equation}
k = \left(\frac{H_1^2 H_P}{V^2} + \frac{\Lambda_{P} H_{P}}{2V} + \frac{\Lambda_{1} H_{1}}{V} +  H_k\right)\hat{e} +\vec{k}
\end{equation}
The vector parts of the one-forms are all manifestly finite in this limit because they contain no inverse powers of V.  The inverses of $Z_1$ and $Z_P$ are also finite.  Examining the metric, all components are finite except for possible poles appearing in front of the $(\hat{e})^2$-term:
\begin{eqnarray}
ds^2 &\sim& \left(-\frac{k_0^2}{Z_1Z_P}+\frac{Z_P}{Z_1}\left[-\frac{k_0}{Z_P}+\frac{H_P}{V}\right]^2+\frac{Z_1}{V}\right)(\hat{e})^2 +...
\end{eqnarray}
However, expanding out these terms we find that the poles at order $V^{-2}$ and $V^{-1}$ identically cancel, thus proving the metric to be finite in this limit. 

Similarly for the B-field, the only terms that contain possible poles are those with one leg along $\hat{e}$:
\begin{eqnarray}
B_2 &\sim&  (dy+dt+\vec{\omega}_P)\wedge \left(\frac{k_0}{Z_1}-\frac{H_1}{V}\right) \hat{e} +...
\end{eqnarray}
but again the poles in this expression cancel identically.  Thus it has been explicitly shown that the solution is well-behaved near any points where $V$ vanishes.

\newsection{Properties of the solution}
\subsection{Mass, Charge and Angular momentum}
Now that the solution has been found to be free of singularities, it is illuminating to study its mass, charges and angular momentum in four dimensions.  Recall that, in six dimensions, the action is
\begin{equation}
S_{6} = \frac{1}{16\pi G_6}\int{d^6x \sqrt{-g} \left(R - \frac{1}{12} G^2\right)}
\end{equation}
To reduce this action to four dimensions in the simplest possible way \cite{maharana}, define
\begin{equation}
ds_6^2 = e^{-\phi_2-\frac{2}{3}\phi_1}d\hat{s}_4^2 + e^{2\phi_2-\frac{2}{3}\phi_1}\left(dz+\hat{\mathcal{A}}_2\right)^2 + e^{2\phi_1}\left(dy+\hat{\mathcal{A}}_1 + \hat{\mathcal{A}}_z dz\right)^2
\end{equation}
and break up the three-form field strength as
\begin{equation}
G = \hat{H}_{(3)}+\hat{G}_{(2)}\wedge(dz+\hat{\mathcal{A}}_2) + \left(\hat{H}_{(2)} + \hat{H}_{(1)}\wedge(dz+\hat{\mathcal{A}}_2)\right)\wedge \left(dy+\hat{\mathcal{A}}_1 + \hat{\mathcal{A}}_z dz\right)
\end{equation}
where all four-dimensional objects are indicated with a hat.  Then the four-dimensional action takes the form
\begin{eqnarray}
S_4 &=& \frac{1}{16\pi G_4}\int d^4x \sqrt{-\hat{g}} \left(\hat{R}_4 -\frac{4}{3} (\partial \phi_1)^2 - \frac{3}{2}(\partial \phi_2)^2 -\frac{1}{2}e^{\frac{8}{3}\phi_1 - 2\phi_2}\hat{F}_{(1)}^2 -\frac{1}{2}e^{-\frac{4}{3}\phi_1 - 2\phi_2}\hat{H}_{(1)}^2 \right. \nonumber \\ &&\left. -\frac{1}{4} e^{3\phi_2} (d\hat{\mathcal{A}}_2)^2 -\frac{1}{4}e^{\frac{8}{3}\phi_1+\phi_2}\hat{F}_{(2)}^2 -\frac{1}{4}e^{-\frac{4}{3}\phi_1+\phi_2}\hat{H}_{(2)}^2 -\frac{1}{4}e^{\frac{4}{3}\phi_1-\phi_2}\hat{G}_{(2)}^2 -\frac{1}{12} e^{\frac{4}{3}\phi_1+2\phi_2}\hat{H}_{(3)}^2 \right) \nonumber 
\end{eqnarray}
where $\hat{F}_{(2)} \equiv d\hat{\mathcal{A}}_1 - d\mathcal{A}_z\wedge\hat{\mathcal{A}}_2$ and $\hat{F}_{(1)} \equiv d\hat{\mathcal{A}}_z$.  For the solution under consideration, the four-dimensional metric is
\begin{equation}
d\hat{s}_4^2 = -\frac{1}{h}(dt+\vec{k})^2 + h(dr^2+r^2d\Omega_2^2)\label{4dMetric}
\end{equation}
where the function $h$ is
\be
h=\sqrt{Z_1^2 Z_P V - k_0^2V^2}\label{h}
\ee
and the two scalars $\phi_{1}$ and $\phi_{2}$ are
\be
e^{2\phi_1} = \frac{Z_P}{Z_1},\  e^{2\phi_2} =\frac{h^2}{Z_1 Z_P V^2} \left(\frac{Z_P}{Z_1}\right)^{\frac{1}{3}}
\ee
while the four gauge field strengths are
\begin{eqnarray}
d\hat{\mathcal{A}}_2 &=& d\chi - \frac{k_0 V^2}{h^2}d\vec{k} -d\left(\frac{k_0V^2}{h^2}\right)\wedge (dt+\vec{k}) \\
\hat{F}_{(2)} &=& d\vec{\omega}_P +\frac{H_P}{V} d\chi -\frac{1}{Z_P} (d\vec{k} + k_0 d\chi) \\ && + \left(\frac{dZ_P}{Z_P^2}\left[1+\frac{k_0^2V^2}{h^2}\right] + \frac{k_0V^2}{h^2}\left[d\left(\frac{H_P}{V}\right)-\frac{1}{Z_P}dk_0\right]\right) \wedge (dt+\vec{k}) \nonumber \\
\hat{H}_{(2)} &=& d\vec{\omega}_1 +\frac{H_1}{V} d\chi -\frac{1}{Z_1} (d\vec{k} + k_0 d\chi)  \\ && + \left(\frac{dZ_1}{Z_1^2}\left[1+\frac{k_0^2V^2}{h^2}\right] + \frac{k_0V^2}{h^2}\left[d\left(\frac{H_1}{V}\right)-\frac{1}{Z_1}dk_0\right]\right) \wedge (dt+\vec{k}) \nonumber \\
\hat{G}_{(2)} &=& *_3 dZ_1 +\frac{k_0}{Z_1}\left(d\vec{\omega}_P+\frac{H_P}{V}d\chi\right) +(dt+\vec{k})\wedge \left(\frac{1}{Z_1}d\left[\frac{H_P}{V}\right] + \frac{1}{Z_P} d\left[\frac{H_1}{V}\right] - \frac{1}{Z_1Z_P}dk_0\right) \nonumber \\ && +\frac{k_0}{Z_1Z_P} \left(Z_1 d\vec{\omega}_1-d\vec{k} + \left[Z_1\frac{H_1} {V}-k_0\right]d\chi\right)
\end{eqnarray}
It will not be necessary to write down $\hat{H}_{(3)}$, $\hat{H}_{(1)}$ or $\hat{F}_{(1)}$ explicitly as they will not be used in the following. They can be easily found from the the ansatz given above.  Note also that the gravitational coupling has been defined as
\begin{equation}
\frac {1}{16\pi G_4} \equiv \frac{1}{2\pi g_s^2 \ell_s^2}\left(\frac{R_y R_z}{\ell_s^2}\right)
\end{equation}
By expanding the function $h$ near infinity, it is easy to deduce that the four-dimensional mass of the current solution is 
\begin{equation}
M = \frac{1}{4G_4} \left(Q_K + 2 Q_1 + \frac{4n(n+1)Q_1^2Q_K}{m^2R_y^2}\right)
\end{equation}
After the identification of the last term above as the momentum charge of the solution, the above relation shows that the mass saturates the BPS bound for the given charges. The angular momentum induced by the $\vec{k}$ term in the metric is
\begin{equation}
J^{12} = -\frac{Q_1^2 Q_K}{2 G_4 mR_y} 
\end{equation}
It is straightforward to write down the charges defined by the above field strengths.  Denoting the physical charges with a tilde, the result of these calculations are, respectively:
\begin{eqnarray}
\tilde{Q}_{K} &=& \frac{Q_K}{4 G_4}, \ 
\tilde{Q}_{P} = \frac{1}{4G_4}\frac{4n (n+1) Q_1^2 Q_K}{m^2 R_y^2} \\
\tilde{Q}_{D1} &=& \frac{Q_1}{4 G_4},\ 
\tilde{Q}_{D5} = \frac{Q_1}{4 G_4}
\end{eqnarray}
We note that if $n=0$ or $-1$ the momentum vanishes. This is the expected behaviour as has been seen in the D1-D5-P case in five dimensions. In the next section the three-charge case will be considered in more detail.

\subsection{Three-charge limit}
The three-charge D1-D5-P solution has the property that if $n=0$ or $n=-1$ the momentum charge vanishes. The four-charge D1-D5-P-KK6 solution has an analogous property in the sense that when $n=0$ or $-1$ the momentum vanishes and the solution reduces to the one found by Bena and Kraus in~\cite{bk} as will be  demonstrated in this section. Start by considering the  $n=0$ case. Substituting for $n$ in Eqs. (\ref{beginsoln})-(\ref{endsoln}) one finds the following nontrivial pieces
\be
V = 1+ \frac{Q_K }{r},\ H_{k} =-\frac{  Q_{1}^2 }{m  Q_{K} R_{y} } \left(1 +   \frac{Q_{K} }{ r_{c}}\right),\ \Lambda_{1} = 1 +  \frac{Q_{1}}{r_{c}},\ H_{P} =   \frac{ 2 Q_{1}^2  }{m Q_{K} R_{y} }  + \frac{m R_{y}}{2} \left( \frac{1}{r } - \frac{1}{r_{c}} \right) \label{n01}
\ee
\be
k = - \frac{Q_{1}^2 Q_{K} }{m c R_{y} } \left( 1- \frac{r+c}{r_{c} } \right) \left( 1+ \cos\theta \right) d\phi + \left( H_{k} + \frac{H_{P}}{2V} \right) \hat{e} \label{n02}
\ee
\be
\omega_{P} =\frac{1}{2} m R_{y} \left[ \frac{ (r - r_{c} ) \cos\theta  + c }{ r_{c} } \right] d\phi + \frac{H_{P}}{V} \hat{e},\ \omega_{1} = 0,\  \hat{e} = dz + Q_{K}  \cos\theta d\phi \label{n03}
\ee
and
\be
c =  \frac{ 4 Q_{1}^2 Q_{K}    }{ m^2 R_{y}^2 - 4   Q_{1}^2 }
\ee
Also, $H_{1}=0$ and $\Lambda_{P}=1$. This implies that $Z_{P}=1$. The relation between $c$ and $R_{y}$ is easily inverted to yield
\be
R_{y}= \frac{2 Q_{1} }{m} \sqrt{ 1+ \frac{Q_{K}}{c} }
\ee
It can be checked that the above functions and one-forms yield the three-charge solution found in~\cite{bk}. It was shown there that for $m=1$ the solution is completely free of horizons and singularities and for $m>1$ it has acceptable $\field{Z}_{m}$ orbifold singularities. This complements our singularity analysis for $n\neq 0,-1$ above. Turn now to the $n=-1$ case. In this case it is convenient to perform the following change of coordinates
\be
r' \cos\theta' = -r\cos\theta - c,\ r' \sin\theta' = r \sin\theta,\ \phi'= -\phi \label{rcnew}
\ee
This transformation effectively interchanges the roles of the points $r_{c}$ and $r$\footnote{Note that the above change of coordinates is similar to the one in Eq. (\ref{rcorigin}). The extra minus signs in Eq. (\ref{rcnew}) are necessary to ensure that not only does the origin in the new coordinate system is at the (old) $r_{c}=0$ but also that the point $r=0$ lies in the new coordinates at $r'=c,\theta'=\pi$ rather than $r'=c,\theta'=0$ as would be the case if we used (\ref{rcorigin}). Finally, to preserve orientation it is necessary to flip the sign of $\phi$ also.}. More precisely one finds $r \rightarrow \sqrt{r'^2 + c^2 + 2 c r' \cos\theta'}\equiv r'_{c}$ and $r_{c} \rightarrow r'$. Thus when $n=-1$ the function $V$ transforms as
\be
V= 1+ \frac{Q_{K}}{r_{c}} \rightarrow 1+ \frac{Q_{K}}{r'}
\ee
The other functions after the coordinate transformation are
\be
H_{k} =\frac{  Q_{1}^2 }{m  Q_{K} R_{y} } \left(1 +   \frac{Q_{K} }{ r'_{c}}\right),\ \Lambda_{1} = 1 +  \frac{Q_{1}}{r'_{c}},\ H_{P} =  -\frac{ 2 Q_{1}^2  }{m Q_{K} R_{y} }  - \frac{m R_{y}}{2} \left( \frac{1}{r' } - \frac{1}{r'_{c}} \right)
\ee
\be
k =  \frac{Q_{1}^2 Q_{K} }{m c R_{y} } \left( 1- \frac{r'+c}{r'_{c} } \right) \left( 1+ \cos\theta' \right) d\phi' + \left( H_{k} + \frac{H_{P}}{2V} \right) \hat{e}
\ee
\be
\omega_{P} = -\frac{1}{2} m R_{y} \left[ \frac{ (r' - r'_{c} ) \cos\theta'  + c }{ r'_{c} } \right] d\phi' + \frac{H_{P}}{V} \hat{e},\ \omega_1=0,\ \hat{e} = dz + Q_{K}  \cos\theta' d\phi'
\ee
Comparing the above functions with the ones appearing in the $n=0$ case as given in Eqs. (\ref{n01})-(\ref{n03}) the functions $Z_{1}$ and $Z_{P}$ are the same as before and the one-forms $k$ and $\omega_{P}$ differ by a sign. This latter sign can be removed by further defining 
\be
t' = -t,\  y'=-y
\ee
Then it is clear that the metric and the gauge field for the $n=-1$ case in the coordinate system $(t',y',r',\theta',\phi',z)$ are the same as the ones in $n=0$ case. Therefore the singularity analysis of \cite{bk} applies in this case as well and it is safe to conclude that the metric is completely smooth for $m=1$ and has a $\field{Z}_m$ orbifold singularity for $m>1$. 

\subsection{Absence of Horizons}

In order to check for horizons one needs to start with the four dimensional solution. The dimensional reduction has been performed earlier and the four dimensional metric was written in Eq. (\ref{4dMetric}). The signal of a horizon would be anywhere that $g^{rr}$ vanishes.  From (\ref{4dMetric}) and (\ref{h}) it is clear that
\be
g^{rr} = \frac{1}{h} = \frac{1}{\sqrt{Z_1^2 Z_P V - k_0^2 V^2}}
\ee
Expanding out this expression in terms of harmonic functions (\ref{beginsoln}-\ref{endsoln}), it takes the form
\be
g^{rr} = \frac{r r_c}{\sqrt{P(r,r_c)}}
\ee
where $P(r,r_c)$ is a polynomial in $r$ and $r_c$.  So $g^{rr}$ vanishes only in two places: $r=0$, $r_c=0$, but it has already been shown that the complete geometry is regular at both these places.  Thus there are no horizons in the geometry.

\subsection{Near-horizon limit}

In order to identify the states constructed above in the dual CFT it is useful to understand its ``near-horizon'' limit\footnote{This terminology is slightly misleading in this case because the solutions under consideration do not have horizons or curvature singularities. ``Near-core'' might be more appropriate.}. This is implemented by taking $\alpha' \rightarrow 0$ and keeping the following combinations fixed: $r \alpha'^{-\frac{3}{2}}$ and $R_{y} \alpha'^{\frac{3}{4}} $. To compute the metric in this limit it is convenient to start by taking the limit of the six harmonic functions and the one forms which specify the metric. It is easily seen from Eqs. (\ref{beginsoln})-(\ref{endsoln}) that this scaling has the effect of eliminating the constant terms in $V,H_{k},H_{P}$ and $\Lambda_{1}$. Furthermore, the relation between $c$ and $R_{y}$ is now given by
\be
c\approx \frac{  a^2  }{ Q_{K} }
\ee
where
\be
a\equiv \frac{2 Q_{1}  Q_{K} }{ m R_{y} }
\ee
Also define the following change of coordinates
\be
Q_{K} r = \rho^2 + a^2 \sin^2\tilde{\theta},\ Q_{K} r \cos^2\frac{\theta}{2} = \rho^2 \cos^2\tilde{\theta},\  z= Q_{K} ( \tilde{\psi} - \tilde{\phi} ), \ \phi = \tilde{\phi} + \tilde{\psi}
\ee
the six dimensional metric can be expressed as
\bea
ds^2 &\approx & - \frac{f}{Q_{1} Q_{K} } (dt^2 - dy^2) + \frac{  n (n+1) a^2 }{Q_{1} Q_{K} } (dt+dy)^2+ 4 a \left( \sin^2\tilde{\theta} d\tilde{\phi} + \cos^2\tilde{\theta} d\tilde{\psi} \right) dy  \nonumber \\
& &  - 4 a (dt+dy) \left[ (n+1)  \sin^2\tilde{\theta} d\tilde{\phi} -n \cos^2\tilde{\theta} d\tilde{\psi} \right]  + \frac{ 4 Q_{1}  Q_{K} }{ \rho^2 + a^2 } d\rho^2 + 4 Q_{1 } Q_{K} d\tilde{\theta}^2 \nonumber \\
 & & + 4 Q_{1} Q_{K} \sin^2\tilde{\theta} d\tilde{\phi}^2 + 4 Q_{1}  Q_{K} \cos^2\tilde{\theta} d\tilde{\psi}^2
\eea
where
\be
f= \rho^2 +  (n+1) a^2 \cos^2\tilde{\theta}  - n a^2 \sin^2\tilde{\theta} 
\ee
This metric can be diagonalised by completing squares on $d\tilde{\psi}$ and $d\tilde{\phi}$. Defining
\be
\bar{\psi} = \tilde{\psi} + n \frac{t}{m R_{y} } + (n+1) \frac{y}{m R_{y}}, \ \ \bar{\phi} = \tilde{\phi} - (n+1) \frac{t}{m R_{y} }  - n \frac{y}{m R_{y}}
\ee
\be
\tilde{t} = \frac{t}{m R_{y} },\ \tilde{\chi} = \frac{y}{m R_{y} },\ \tilde{\rho} = \frac{\rho}{a} 
\ee
brings the metric to locally $AdS_{3}\times S^{3}$
\be
ds^2 \approx L^2 \left[ -(\tilde{\rho}^2 +1 ) d\tilde{t}^2 + \tilde{\rho}^2 d\tilde{\chi}^2 + \frac{ d\tilde{\rho}^2 }{ \tilde{\rho}^2 +1} + d\tilde{\theta}^ 2+ \cos^2\tilde{\theta} d\bar{\psi}^2 + \sin^2\tilde{\theta} d\bar{\phi}^2 \right] 
\ee
with the scale
\be
L^2 = 4 Q_{1} Q_{K}
\ee
 
The identifications on the final coordinates $\bar\psi$, $\bar\phi$ and $\chi$
are 
\be
(\bar{\psi},\bar{\psi},\tilde{\chi})\cong (\bar{\psi},\bar{\psi},\tilde{\chi})+{2\pi n_1\over N_K}\,(1,-1,0)+2\pi n_2\,(1,0,0)+{2\pi n_3\over m}\,(n+1,-n,1)
\ee
For $m=1$ this defines an orbifold $AdS_{3}\times(S^{3}/\mathbb{Z}_{N_K})$, 
corresponding to $N_K$ coinciding KK monopoles. For $m>1$ the 6D geometry
is further orbifolded by a $\mathbb{Z}_m$ group, whose action mixes the 
$S^3$ and $AdS_3$ coordinates, and which corresponds to a conical defect
of order $m$.
  
\newsection{Discussion}

In this paper an axially symmetric, rotating four-dimensional solution carrying D1,D5, KK6 and momentum charges has been constructed. The solution exhibits complete regularity and has a near-core geometry of $AdS_{3}\times S^{3}$. It has been constructed within the framework of six dimensional minimal supergravity. As a result the D1 and D5 charges are equal and the dilaton is trivial. This choice simplifies the computation but one could in principle generalise our solution to $Q_{1} \neq Q_{5}$ by using the formalism of~\cite{bw}. As is shown in Appendix A, the geometry constructed here cannot be derived from the standard rotating non-extremal solution~\cite{cvetic4D}, just as the rotating BPS black hole~\cite{blackhole4D} cannot be derived from this class. However, the construction of the black hole suggests that if one could embed the non-supersymmetric black ring~\cite{blackring} in Taub-NUT space to generate a new non-extremal family  of solutions, it might be possible to reproduce the smooth D1-D5-KK6-P solution by taking the BPS limit in this class.

The most important question regarding this solution (and the one with zero momentum) is whether it describes a bound state of the branes. This question is important if one is to interpret the solution as being a microstate of the D1-D5-KK6 system. Recall that in the current construction a known D1-D5-P microstate was the starting point, and then it was embedded it in a KK background. Thus it is reasonable to conjecture that the D-branes and momentum are bound to each other. It would be good to know if the KK monopoles are bound to the other charges as well. A necessary condition for this to happen is that the solution be invariant under string dualities which interchange two or more of the charges. In particular one could perform the sequence $S T_{z} T_{1} S$ which would interchange the D5 branes with the KK monopoles. Here $T_{z}$ represents a T-duality along the fiber direction of the KK monopole and $T_{1}$ is a duality along any one of the directions of the internal $\field{T}^4$. Unfortunately if one applies this sequence to the current solution with $Q_{1}=Q_{5}$ one moves out of the domain of minimal supergravity. In order to perform this check the solution needs to be generalised to allow independent $Q_{1}$ and $Q_{5}$. However the solution constructed in~\cite{bk} is amenable to this check and it would be very interesting to know whether it is symmetric under this duality transformation.
 
Most of the properties of the KK6-D5-D1-P solution are qualitatively similar to the ones found for the D1-D5-P system. A general feature of the singularity resolution in both  systems is the mixing between the base metric and the $y$ circle. This mixing comes about because of the behaviour of $\omega_{P}$ at the poles of $V$. At the poles, $\omega_{P}$ approaches a non-zero, constant one-form proportional to $m R_{y}$. In the D1-D5 system, the positive integer $m$ parametrizes the multiplicity of the winding of the effective string. This interpretation excludes the possibility of $m=0$ and hence the possibility of $\omega_{P}$ vanishing at the poles. One would expect a similar interpretation of $m$ to be true in the D1-D5-KK6 system and thereby excluding the $m=0$ case. 

It was noted in~\cite{bk} that in the D1-D5-KK6 solution the gauge field which gave rise to the KK monopole charge in four dimensions was also charged electrically. This feature persists when momentum is added to the system but the value of this extra charge changes. In the current case, one finds
\be
Q_{KKe} = \frac{2m(2n+1) R_{y} Q_{1}^2 }{m^2 R_{y}^2 + 4 n (n+1) Q_{1}^2 }
\ee
For $n=0$ it coincides with the value noted in~\cite{bk}. The interesting aspect of this expression is its dependence on $n$.  There is an extra piece in the denominator due to the momentum. However in the near-horizon region where $R_{y} \gg Q_{1}$ this piece can be ignored and one gets
\be
\left. Q_{KKe}\right|_{n.h.} =  \frac{2(2n+1) Q_{1}^2 }{m R_{y} }
\ee
The remaining dependence on $n$ is significant from the CFT point of view. To understand the possible role played by this charge in the CFT, first consider the angular momentum of the system. The angular momentum for the D1-D5-KK6-P system is given by
\be
J = -\frac{Q_1^2 Q_K}{2 m R_{y} G_4 } 
\ee
Rather surprisingly this expression does not depend on $n$. In other words, the value of the angular momentum is fixed completely by the D1, D5 branes and the KK monopoles and is independent of the momentum. To get some physical insight into this behavior it is useful to express the two relations above in terms of dimensionless quantities
\be
J= -\frac{1}{2m} N_{1}^2 N_{K},\ \ N_{e} \equiv \frac{Q_{KKe}R_y R_z^2}{g_s^2\ell_s^4} =  \frac{2n+1}{2m} N_{1}^2 
\ee
A related issue concerns the CFT interpretation of the integer $n$. In the D1-D5 system, one encounters an analogous integer parameter and there it was related to the property of spectral flow in the CFT. 

The full D1-D5 CFT is a sigma model with target space a deformation of $(T^{4})^{N}/S_{N}$ where $N=N_{1}N_{5}$ and central charge, $c=6 N_{1}N_{5}$. For current purposes it suffices to restrict attention to just one copy of the CFT with target space $T^{4}$ and $c=6$. Consider one side of the CFT with NS boundary conditions on the fermions. The NS vacuum has $h=j=0$. Spectral flow~\cite{ss} maps NS sector states to R sector states and vice versa under which the conformal dimension and spin change as
\be
h' = h- \alpha q + \alpha^2 \frac{c}{24}, \ \ q' = q- \alpha \frac{c}{12},\ \ \alpha\in \field{Z}
\ee
If $\alpha$ is odd one goes from NS sector to the R sector. The field theory on the D1-D5 branes is in the Ramond sector and the asymptotically flat geometries describe gravity duals of R sector states. Spectral flow can be performed independently on the left and the right side. To get a state which preserves some part of the supersymmetry of the vacuum we perform spectral flow with $\alpha_R=1$ on one of the sectors, say, the right movers. On the other side one sets $\alpha_L=2n+1$ such that the NS ground state maps to an excited R state. The resulting state after spectral flowing the NS ground state with $h_{NS} =j_{NS}=\bar{h}_{NS}=\bar{j}_{NS}=0$ has conformal dimension and spin
\be
h_{R}= \frac{1}{4} (2n+1)^2 N_{1} N_{5}, \ j_{R}= -\frac{1}{2} (2n+1)  N_{1} N_{5},\  \bar{h}_{R} = \frac{1}{4} N_{1} N_{5}, \  \bar{j}_{R} = -\frac{1}{2} N_{1} N_{5}
\ee
The state therefore carries momentum given by $P= n(n+1) N_{1} N_{5}$. 

The story thus far does not inculde KK monopoles. The presence of KK monopoles destroys half of the supersymmetry (i.e. a $(0,4)$ CFT with only $su(2)_{R}$ R-symmetry algebra) and modifies the central charge to $c=6 N_{1} N_{5} N_{K}$.  Since spectral flow acts on each of the sides independently, the right movers (i.e. the supersymmetric side) still admit spectral flow. However it is not clear whether there is anything like a ``spectral flow'' on the non-supersymmetric side of the CFT. If the supergravity state constructed in this paper corresponds to a microstate of the D1-D5-KK6 CFT it would seem that the non-supersymmetric sector of the CFT admits a spectral flow-like transformation with the role of R-charge being played by $N_{e}N_{K}$.  This conjecture however, depends crucially on whether the states that have been constructed describe bound states of the branes or not. Assuming that the geometry is dual to a microstate in the CFT, the observed charges and angular momentum can be explained by starting with the ground state in the NS sector and spectral flowing once on the supersymmetric side while performing the ``non-supersymmetric spectral flow'' $2n+1$ times. The angular momentum of the gravity solution corresponds to the R-charge from the supersymmetric side and the momentum is proportional to $h-\bar{h}$ as usual. It would be good to understand these issues in more detail because they may shed some light on the structure of the CFT underlying D1-D5-KK6.

\section*{Acknowledgements}

The authors would like to thank Samir Mathur and Yogesh Srivastava for several discussions, and to the organizers of the Workshop on Gravitational Aspects of String Theory held at the Fields Institute, where this work was begun. GP would like to thank Omid Saremi. AS, GP and AWP acknowledge support from the National Science and Engineering Research Council of Canada (NSERC) and the Canadian Institute for Advanced Research (CIAR); 
SG was supported by the Istituto Nazionale di Fisica Nucleare (INFN).

\section*{Appendix A: Non-extremal to Extremal?}
\begin{appendix}
\renewcommand{\theequation}{A.\arabic{equation}}
\setcounter{equation}{0}
\renewcommand{\thesubsection}{A.\arabic{subsection}}
\setcounter{subsection}{0}

A standard method of constructing supersymmetric solutions in supergravity is by taking the supersymmetric limit of a non-extremal solution. The non-extremal solution in turn can be constructed by using the solution generating techniques of~\cite{sen}. The idea of this method is to start with some ``seed'' solution which is an exact solution of Einstein gravity and add charges by applying the stringy dualities and boosts. In practice one typically starts with Kerr/Myers-Perry~\cite{myersperry} solution in some dimension (depending on isometries required in the final solution) and lifting it to ten or eleven dimensions by adding requisite number of flat toroidal directions before performing boosts and T-dualities along the compact directions accompanied by some S-dualities. In this manner one can construct classes of non-extremal black holes carrying angular momentum and up to three charges in five dimensions or four charges in four dimensions. To construct a supersymmetric solution, one can take a BPS limit of the non-extremal solutions by sending the excess mass to zero while keeping the physical quantities fixed. The simplest nontrivial class has the Kerr/Myers-Perry solution in five dimensions with  two angular momenta as the seed which after boosts and dualities yields a non-extremal solution carrying three charges and two angular momenta. The most common duality frame employed to write the solution is the D1-D5-P system where the two branes and the momentum share a common direction, say $y$ and the D5 branes are also extended along a four torus. The radius of the $y$ circle is taken to be much larger than the other four compact directions. For this reason the metric is most conveniently expressed in six dimensions.

A well known example of this approach to producing supersymmetric solutions is the BMPV black hole~\cite{bmpv}. However the method has been useful not only for objects with a horizon but also for smooth microstates. If one carefully takes the BPS limit of the same class of solutions which yielded the BMPV solution one finds a three charge microstate, as was demonstrated in~\cite{gms1}. A slightly unconventional feature of the limiting procedure was that the angular momentum parameters were divergent as the non-extremality $M$ was sent to zero. One can easily deduce the limiting behavior of these parameters by considering the near horizon geometry and demanding smoothness. Let the D1 and D5 charges be denoted by $Q_{1}$ and $Q_5$ respectively and the radius of the $y$ circle be $R_{y}$. The near-extremal near horizon geometry was found in~\cite{cvetic5Dnh} and is the BTZ black hole times a three sphere. The BTZ part of the geometry is
\be
ds^2_{n.h.} = -N^2 d\tau^2 + N^{-2} d\rho^2 + \rho^2 \left( d\sigma - N_{\sigma} d\tau \right)^2 
\ee
where
\be
N^2= \frac{\rho^2}{\lambda^2} - M_{3}  + \frac{16 G_{3}^2 J_{3}^2}{\rho^2}, \ N_{\sigma} = \frac{4 G_{3} J_{3} }{\rho^2},\ \ \lambda^{4} = Q_{1} Q_{5}
\ee
The mass $M_{3}$ and the  angular momentum $J_{3}$ of the above geometry are
\bea
M_{3} &=&  \frac{R_{y}^2}{\lambda^4 } \left[ (M - a_{1}^2 - a_{2}^2 ) \cosh 2\delta_P + 2 a_{1} a_{2} \sinh2\delta_{P} \right] \label{mass5d}\\
8  G_{3} J_{3}&=& \frac{R_{y}^2}{\lambda^3 } \left[ (M - a_{1}^2 - a_{2}^2 ) \sinh 2\delta_P + 2 a_{1} a_{2} \cosh 2\delta_{P} \right] \label{ang5d}
\eea
The boost parameter $\delta_{P}$ is related to the momentum through
\be
Q_{P} = \frac{M}{2} \sinh 2\delta_{P}
\ee
and in the BPS limit we take $M\rightarrow 0$ while keeping $Q_{P}$ fixed. In order to have a smooth asymptotically flat geometry it is necessary that the near horizon geometry be free of singularities. For the BTZ black hole this is the case if $J_{3}$ is zero and the mass $M_{3}$ is $-1$ in which case it reduces to global $AdS_{3}$.  Assuming these values in the equations above and solving for the $M\rightarrow 0$ behavior of $a_{1}$ and $a_{2}$ one finds a consistent solution
\bea
a_{1} &=& - \frac{1}{R_{y} } \sqrt{\frac{Q_{1}Q_{5} Q_{P}}{M}}+ \frac{\sqrt{Q_{P} R_{y}^2 + Q_{1} Q_{5} }}{4R_{y}} \sqrt{\frac{M}{Q_{P}}}\\ a_{2} &=& - \frac{1}{R_{y} } \sqrt{\frac{Q_{1}Q_{5} Q_{P}}{M}}- \frac{\sqrt{Q_{P} R_{y}^2 + Q_{1} Q_{5} }}{4R_{y}} \sqrt{\frac{M}{Q_{P}}}
\eea

After substituting for the momentum charge expected from CFT considerations one can verify that this behaviour agrees with the one that was derived in~\cite{gms1} by considering the full asymptotically flat geometry. Note that as $M\rightarrow 0$, $a_{1}$ and $ a_{2}$ are divergent. However, this divergence can be completely removed by a shift in the radial coordinate and the resulting metric is smooth and horizon free.  

Given this success in five dimensions, it is natural to ask whether one can construct microstates in four dimensions by starting with the rotating four-charge non-extremal solution constructed in~\cite{cvetic4D} and taking the extremal limit. It will be shown in the following this is not possible. The simplest way to show this is to use the near-horizon near-extremal geometry of these solutions~\cite{cvetic4D} which is again a BTZ black hole but this time tensored with a two sphere. We work in the KK6-D5-D1-P duality frame. The argument is very similar to the one above for the five-dimensional case. The mass and angular momentum of the BTZ are given by ($\lambda^3= Q_{1} Q_{5} Q_{K}$)
\bea
M_{3} &=&  \frac{R_{y}^2}{\lambda^3 } \left[ ( 4M - \frac{2a^2}{M}  ) \cosh 2\delta_P + \frac{2 a^2}{M} \sinh2\delta_{P} \right] \label{mass4d}\\
8  G_{3} J_{3}&=& \frac{R_{y}^2}{\lambda^2 } \left[ (4M- \frac{2 a^2}{M}) \sinh 2\delta_P + \frac{2 a^2}{M} \cosh 2\delta_{P} \right] \label{ang4d}
\eea
As before the boost parameter $\delta_{P}$ is related to the momentum $G_{4} Q_{P} = M \sinh2\delta_{P}/4$ and $a$ is related to the angular momentum of the asymptotically flat geometry. One can already anticipate the nature of the problem one will encounter when trying to impose the condition that the near-horizon geometry be smooth in the $M\rightarrow 0$ limit. There are two equations and only one variable whose limiting behaviour is unknown, i.e. $a$. This hurdle can in principle be avoided by allowing $Q_{P}$ to be fixed by the above equations. However one finds that as $\delta_{P}\rightarrow \infty$ the two equations can be satisfied simultaneously with $M_{3}=-1$ and $J_{3}=0$ only if $Q_{P} <0$, which contradicts our assumption that $\delta_{P}>0$. In other words there is no choice of parameters such that global $AdS_{3}$ is recovered for the near-horizon geometry, in direct contrast to the case of three charges in five dimensions.

This discussion has shown that the non-extremal approach to constructing four dimensional microstates would not succeed. However, dramatic developments in the understanding of supersymmetric solutions of various supergravities, for example in\cite{gmr},\cite{bw} will allow one to attack this problem in a more direct manner. This method has already been applied in~\cite{bk} to construct KK6-D5-D1 microstates in four dimensions and in the body of the paper such developments are used to add momentum to this system. 

\renewcommand{\theequation}{B.\arabic{equation}}
\setcounter{equation}{0}
\renewcommand{\thesubsection}{B.\arabic{subsection}}
\setcounter{subsection}{0}
\section*{Appendix B: Solving for one forms}
There are two kinds of equations which are encountered for the one-forms appearing in our problem. The first is the equation which determines the three-dimensional parts of $\omega_{1}, \omega_{P}$ and the one form $\chi$ (which is purely three-dimensional) . Denote the vector part of any one of these forms by $\vec{\alpha}$. Then the equation satisfied by $\vec{\alpha}$ is of the form
\be
\ *_{3} d \vec{\alpha} = d J
\ee
Here $J$ is some harmonic function in three dimensions which discrete sources. Restrict oneself to the case in which $J$ has only two point sources located respectively at the origin and at $\theta=\pi, r=c$. With these assumptions write
\be
J=  \frac{a}{r} + \frac{b}{\sqrt{r^2 + c^2 + 2 c r \cos\theta} }
\ee
By linearity the dual of the two pieces can be calculated separately and then added. For the piece with the source at the origin the result is well known
\be
\ *_{3} d \left( \cos\theta d\phi \right) = d \left(\frac{1}{r} \right)
\ee
The simplest way to find the corresponding result for a source at an arbitrary point is to use the translation symmetry of $\field{R}^3$. Though this is easy to do for an arbitrary point the resulting expression in spherical coordinates is much simpler if the point is on the axis ($\theta=0,\pi$). In the spherical coordinate system this translation is acheived by defining
\be
(r,\theta) \rightarrow (r',\theta'): r \cos\theta = r' \cos\theta'+c, \ r\sin\theta= r'\sin\theta'
\ee
Computing the expression for $\cos\theta$ from the above in terms of the new coordinates and dropping the primes the dual of the second piece in $J$ is found. Combining the two results:
\be
\vec{\alpha} = \left( a \cos\theta  + b \frac{ r \cos\theta + c }{ \sqrt{ r^2 + c^2 + 2 c r \cos\theta } } \right) d\phi
\ee
Applying this result to find $\chi$, the one-form appearing in the base metric and with the source $J= V$ is
\be
V= 1 + \frac{Q_{K}}{\tgamma_{1} + \tgamma_{2} } \left(  \frac{\tgamma_{2}}{r } + \frac{ \tgamma_{2} }{ r_{c} } \right),\ \chi = \frac{Q_{K}}{\tgamma_{1} + \tgamma_{2} } \left( \tgamma_{2} \cos\theta + \tgamma_{1}  \frac{ r \cos\theta + c}{r_{c} } \right)  \label{chieqn}
\ee
where we have named the frequently appearing combination $r_{c} = \sqrt{ r^2 + c^2 + 2 c r \cos\theta}$. The expressions for the three-dimensional parts of $\omega_{1}$ and $\omega_{P}$ are found similarly with source $J$ being $ -H_{1}$ and $-H_{P}$ respectively.

The second kind of dualization needed in the main text is in the equation which determines $\vec{k}$ in terms of the six harmonic functions given in the third line of Eq. (\ref{gmrsol}). The source on the right hand side of this equation can be naturally divided into three pieces each of which is by itself integrable (i.e. the operator $d\*_{3}$ acting on the term is zero). Thus it suffices to consider any one of the three terms. Let $J_{1}$ and $J_{2}$ be two harmonic functions with common two point sources, i.e.  let
\be
J_{i} \equiv  \alpha_{i} + \frac{\beta_{i} }{r} + \frac{\delta_{i} }{r_{c} },\ i=1,2
\ee
Then by a direct computation (note that $\vec{c} = ( 0,0,c)$)
\be
J_{1} d J_{2} - J_{2} d J_{1} = - ( \alpha_{1} \beta_{2} - \alpha_{2} \beta_{1} ) \frac{dr}{r^2} - ( \alpha_{1} \delta_{2} - \alpha_{2} \delta_{1} ) \frac{(\vec{x} + \vec{c} ). d\vec{x} }{r_{c}^3} - ( \beta_{1} \delta_{2} - \beta_{2} \delta_{1} ) \left[ \frac{( \vec{x} + \vec{c} ).d\vec{x} }{r r_{c}^3 } - \frac{dr }{r^2 r_{c} } \right]
\ee
The first and the second terms above can be dualised easily using the result for $\chi$ while the third piece requires more work.  Taking the hodge star of the above and rearranging, one finds:
\bea
*_{3} ( J_{1} d J_{2} - J_{2} d J_{1} ) &=& d\left[ \left\{ ( \alpha_{1} \beta_{2} - \alpha_{2} \beta_{1} ) \cos\theta  + ( \alpha_{1} \delta_{2} - \alpha_{2} \delta_{1} ) \frac{r \cos\theta + c }{r_{c} }  \right. \right. \nonumber \\ & & \ \ \ \ \ \ \ \left. \left.+ ( \beta_{1} \delta_{2} - \beta_{2} \delta_{1} ) \frac{ r_{c}-r - c \cos\theta }{ c r_{c} } \right\}  d\phi \right]
\eea
It is now straightforward to put together the various terms which contribute to the curl of $\vec{k}$. In general,
\be
\vec{k} = \left\{ f_{1} \cos\theta  + f_{2}  \frac{r \cos\theta - c }{r_{c} }  + f_{3} \frac{ r -r_{c} - c \cos\theta }{ c r_{c} } \right\}  d\phi
\ee
with $f_{1},f_{2}$ and $f_{3}$ given in Eq. (\ref{f1})-(\ref{f3}).
\end{appendix}

\end{document}